\documentclass[10pt,journal]{IEEEtranTCOM}

\IEEEoverridecommandlockouts	
\usepackage[dvips]{graphicx}
\usepackage{graphicx}
\usepackage{amssymb}
\usepackage{cite}
\usepackage{subfigure}
\usepackage{multirow}
\usepackage{bm}
\usepackage{amsmath}
\usepackage{amsthm}
\usepackage{stfloats}
\usepackage{float}

\usepackage{subeqnarray}
\usepackage[linesnumbered,ruled]{algorithm2e}
\usepackage{algpseudocode}
\usepackage{color}
\usepackage{cases}
\usepackage{caption2}
\usepackage{url}
\usepackage{array}
\usepackage{enumerate}
\usepackage{tabularx}
\usepackage{multirow}
\usepackage[table,xcdraw]{xcolor}
\usepackage{diagbox}
\usepackage{booktabs}
\usepackage{colortbl}
\makeatletter

\newcommand{\Rmnum}[1]{\expandafter\@slowromancap\romannumeral #1@}
\makeatother
\renewcommand{\baselinestretch}{1}

\theoremstyle{plain}
\newtheorem{theorem}{\textit{Theorem} \textit}
\newtheorem{lemma}{Lemma}

\theoremstyle{definition}

\theoremstyle{remark}
\newtheorem{remark}{\textbf{Remark}\bf }

\begin{document}
\title{ISAC-Enabled Beam Alignment for Terahertz Networks: Scheme Design and Coverage Analysis}
\author{Wenrong Chen, \IEEEmembership{Student Member, IEEE}, Lingxiang Li, Zhi Chen, \IEEEmembership{Senior Member, IEEE},
\\ Yuanwei Liu, \IEEEmembership{Fellow, IEEE}, Boyu Ning, Tony Quek, \IEEEmembership{Fellow, IEEE}
	\thanks{
		Copyright (c) 2024 IEEE. Personal use of this material is permitted. However, permission to use this material for any other purposes must be obtained from the IEEE by sending a request to pubs-permissions@ieee.org.}
	\thanks{
		This work was supported in part by the National Natural Science Foundation of China (NSFC) under Grant U21B2014 and 62271121, and the National Research Foundation, Singapore and Infocomm Media Development Authority under its Future Communications Research \& Development Programme.
		\textit{(Corresponding author: Zhi Chen.)}}
	\thanks{
		\IEEEcompsocthanksitem W. Chen, L. Li, Z. Chen and B. Ning are with National Key Laboratory of Wireless Communications, University of Electronic Science and Technology of China, Chengdu 611731, China (E-mails: wenrongchen@std.uestc.edu.cn; \{chenzhi, lingxiang.li\}@uestc.edu.cn;boydning@outlook.com).
         \IEEEcompsocthanksitem Yuanwei Liu is with the Department of Electrical and Electronic Engineering, The University of Hong Kong (E-mail: yuanwei@hku.hk).	
		\IEEEcompsocthanksitem T. Q. S. Quek is with the Singapore University of Technology and Design, Singapore 487372 (e-mail: tonyquek@sutd.edu.sg).		
	}
}
\maketitle

\begin{abstract}
As a key pillar technology for the future 6G networks, Terahertz (THz) communications can provide high-capacity transmissions, but suffers from severe propagation loss and line-of-sight (LoS) blockage that limits the network coverage.
Narrow beams are required to compensate for the loss, but they in turn bring in beam misalignment challenge and degrade the THz network coverage.
The high sensing resolution of THz signals enables integrated sensing and communications (ISAC) technology to assist the LoS blockage and user mobility-induced beam misalignment, enhancing THz network coverage.
Based on the 5G beam management, we propose a joint synchronization signal block (SSB) and reference signal (RS)-based sensing (JSRS) scheme to assist beam alignment.
JSRS enables a predict-and-prevent procedure that provides early interventions for timely beam switches.
To maximize performance of JSRS, we provide an optimal sensing signal insertion and time-to-frequency allocation to improve the joint range and velocity resolutions.
We derive the coverage probability of the JSRS-enabled network to evaluate its abilities in beam misalignment reduction and coverage enhancement.
The expression also instructs the network density deployment and beamwidth selection.
Numerical results show that the JSRS scheme is effective and highly compatible with the 5G air interface.
Averaged in the tested urban use cases, JSRS achieves near-ideal performance and reduces around 80\%  of beam misalignment, and enhances the coverage probability by about 75\%, compared to the network with 5G-required positioning ability.


\end{abstract}


\IEEEpeerreviewmaketitle

\section{Introduction}
Towards the realization of smart cities, the future 6G foresees the implementation of authentic digital twin representation of the physical world, supporting novel use cases, including but not limited to vehicle-to-everything (V2X) communications, unmanned air vehicles (UAV) and smart factories\cite{liu2022integrated}.
The ultra-wide terahertz (THz) band that can support massive connections with up to terabit per second (Tbps) data rate has come into vision\cite{ian2022terahertz}.

However, THz-band transmissions suffer from limited signal power, high propagation loss, and severe molecular absorption, causing the coverage bottleneck\cite{jornet2011channel}.
Highly directional beams are exploited to enhance the signal power, but they bring in narrow beam alignment challenges\cite{elbir2022terahertz}.
Specifically, for high-frequency communication systems, beam steering is a basic transmission technology to provide beamforming gain along the desired direction, while beam alignment technology searches over the codebook to find the pair of beam steering vectors at the transceivers, which maximizes the received signal-to-noise ratio (SNR).
Narrower beams indicate larger codebook size and unacceptable beam alignment overhead.
Furthermore, the THz link is sensitive to line-of-sight (LoS) blockage, meaning objects larger than several wavelengths can cause beam misalignment and interrupt link connection\cite{ning2023beamforming,wu2020interference}.
Thus, highly directional beams bring about a high probability of beam switches and non-negligible latency for link reconnection, which severely degrades the communication quality, including the coverage probability and throughput\cite{kalamkar2021beam,chen2021mobility}.
It is essential to develop an efficient beam management scheme for THz networks \cite{ning2021terahertz,ning2021unified}.


Traditional communication-based beam management follows the 3GPP specified detect-error-and-correct beam recovery procedures, which is triggered after the beam failures are detected\cite{dahlman20205g}.
Introducing ISAC into the network brings opportunities to predict the upcoming beam switches and thus prevents beam failures, allowing seamless beam switches\cite{sarieddeen2020next}.
THz signals' intrinsic high-resolution sensing ability, opens up unprecedented opportunities to realize integrated sensing and communications (ISAC) in THz networks\cite{han2022thz}.
It allows for joint sensing and communications using a unified waveform and integrated hardware\cite{chen2021terahertz}.
Compared to the traditional sensor-aided solutions, ISAC uses shared resources that can achieve low latency and high spectral efficiency with reduced cost\cite{wild2021joint}.
The state-of-the-art THz band ISAC prototypes have realized millimetre-level imaging resolution, showing its feasibility to assist THz communications\cite{li2022integrated}.
However, from the ISAC prototype to the implementation of sensing-aided THz communications, the following challenges need to be addressed:
1) The design of a sensing-aided beam alignment scheme that is compatible with the current 5G beam management procedure;
2) The flexible sensing signal configuration that efficiently uses the sensing resources in varying networks;
3) Network-level performance analysis that captures the enhancement provided by the sensing assistance.

To tackle the aforementioned challenges, many foundational works have contributed to the link-level ISAC-THz techniques.
Works\cite{zhang2021overview,liu2023seventy,chaccour2021joint,wu2022sensing} focus on sensing information extraction and ISAC waveform design.
For candidate standard signal selection, the standard signals are preferred to be sensing candidates to realize better compatibility with the 5G air interface.
Theoretically, work\cite{zhang2020perceptive} compares the feasibility of using the channel estimation signals, \textit{synchronization signal block} (SSB), \textit{demodulation reference signals} (DMRS), and \textit{positioning reference signals} (PRS) for sensing.
As pilot signals, they shine with intrinsic auto-correlation and anti-noise properties that improve localization accuracy\cite{wei20225g}.
Work\cite{sun2022embedding} develops reference signals (RS)-based angle and time estimation algorithms, and evaluates the superiority of using reference signals over other standard signals for localization in V2X networks.
Work\cite{wei20225g} provides frame structure design for PRS-based sensing signal configuration to perform velocity estimation.
However, the sensing-aided communications scheme that benefits network-level performance has not been well investigated.

Till now, few works have been done to bridge the gap from the link level to the network level, covering areas from the scheme design and sensing signal configuration to performance evaluation.

As surveyed by \cite{lu2024integrated}, BS coordinated ISAC has shown its potential to improve the SNR and the detect coverage, achieving gains in both communication and sensing functions.
Works \cite{sheemar2023full} emphasizes the importance of developing ISAC networking strategies to achieve mutual gains.
Focusing on sensing-aided vehicular networks, work \cite{meng2023uav} shows the future directions in developing network-level performance evaluation frameworks.
Conceptual ideas have been raised, but specific networking strategies still need further discussions.
As a preliminary discussion on the scheme design, our previous work\cite{chen2022enhancing} proposes an SSB-based sensing method to assist the THz/mmWave network beam alignment.
However, its benefits on network coverage have not been analysed.
Considering the cost and benefit of sensing, work \cite{ji2023networking} designs a networking-based ISAC hardware testbed that reveals the resource allocation trade-off between communication and sensing functions.
However, it does not provide an effective scheme to configure the sensing signals, and no sensing-aided communications scheme has been considered.

For network performance analysis, stochastic geometry (SG) is a well-investigated tool that has been widely used to model networks ranging up to THz bands.
Key performance indicators ranging from interference to the coverage probability can be modelled using the basic theories developed in \cite{baccelli2010stochastic,haenggi2009stochastic,baccelli2009stochastic}.
The LoS blockage model and its impact on the network interference are discussed in \cite{petrov2017interference}, and the coverage probability is analysed in\cite{humadi2021coverage,shi2021coverage}.
Recently, SG has also been exploited for ISAC network modelling.
Work\cite{olson2022coverage} develops a mathematical framework that characterizes the coverage probability of ISAC-based networks, but the coverage enhancement by introducing sensing into the system has not been analysed.
Moreover, these works for ISAC are for mmWave and below bands.
For the THz network with ISAC, the unique THz propagation features, LoS blockage and mobility effects, need to be considered.
There still remain blanks for the analysis on the ISAC benefit in THz network coverage.

To solve the beam misalignment challenge in THz networks at the network level, we propose an ISAC-based beam alignment approach that enhances the network coverage, which is in line with the 5G standard.
The main contributions are summarised as follows.

1) We propose a joint SSB and RS-based sensing (JSRS) scheme to reduce the LoS blockage and user mobility-induced beam misalignment in THz networks, which is in line with the 5G beam management procedure.
JSRS exploits the 5G-specified SSBs to detect blockages and the RSs for user tracking.
Different from the traditional detect-and-correct method, sensing-based JSRS enables a predict-and-prevent procedure that provides early interventions for timely beam switches, reducing beam misalignment.


2) Based on the 5G frame structure, we provide an optimal sensing signal pattern design which minimizes beam misalignment with a given number of sensing resources.
Specifically, we optimize the time-to-frequency allocation ratio that minimises the sensing error-induced beam misalignment, and the insert spacings to satisfy the sensing range and resolution requirements.
We reveal the design trade-offs in the time-to-frequency allocation ratio affected by the beamwidth and frequency bands (\textit{see Theorem \ref{ratio_opt}}).

3) From the network level, we then analyse the coverage probability of the ISAC-THz network, using the stochastic geometry.
The expression enables evaluations on the benefit of introducing sensing into THz networks and provides instructions for the ISAC-THz network deployment that achieves higher coverage probability,
including the design of network density, beamwidth, and sensing signal configuration (\textit{see Theorem \ref{theo_pcvp}}).

4) Based on the numerical analyses, we show the effectiveness of the JSRS scheme and its high compatibility with the 5G air interface.
In tested urban V2X use cases, the JSRS scheme averagely reduces the beam misalignment by 80\%, and increases the coverage probability by  75\%, compared to that of the 5G-ability networks.
Besides, the JSRS scheme achieves near-ideal performance, showing that the proposed sensing signal configuration effectively reduces the estimation error.
JSRS exploits the high angular resolution of narrow beams to enhance the distance estimation and turns the disadvantage of sharp beams into benefits, allowing the THz network to use narrower beams to enhance coverage.
%

The remainder of this paper is organized as follows.
Sec.\ref{sec_system} introduces the ISAC-THz network model and the definitions for the coverage probability and beam misalignment.
Sec.\ref{sec_JSRS} proposes the JSRS scheme for THz beam alignment and analyses its feasibility.
Sec.\ref{sec_ability} analyses its sensing ability and the arose problem of sensing resource allocation.
Sec.\ref{sec_optimal_pattern} provides a time-frequency sensing signal pattern design and evaluates the cost of sensing.
The analytical expression for the coverage probability of the ISAC-THz network is given in Sec.\ref{sec_cov_thr}.
Numerical results are presented in Sec.\ref{sec_numerical}.
Conclusions are drawn in Sec.\ref{sec_conclusion}.
The definitions of frequently used symbols are summarized in Table \ref{table.definition}.


\section{System Model and Performance Metrics}\label{sec_system}
This section first introduces the system model of the considered ISAC-enabled THz network with randomly distributed base stations (BSs) and mobile terminals (MTs).
Next, we introduce the definitions of beam misalignment and coverage probability that depict the network coverage ability.
\begin{table}[tb]
	\centering
	\caption{Symbols definition.}
	\newcolumntype{Y}{>{\raggedright\arraybackslash}X}	
\begin{tabular}{r|l}
	\hline
	\rowcolor[HTML]{DEDEDE}
	\textbf{Symbol}                                      & \textbf{Definition}                                          \\ \hline
	\rowcolor[HTML]{FFFFFF}
	$r_{\rm{B}}$                                         & Node radius                                                  \\
	\rowcolor[HTML]{FFFFFF}
	$\lambda_B,\lambda_M,\lambda_S$                      & Densities of BSs, MTs, and blockers                          \\
	\rowcolor[HTML]{FFFFFF}
	$n_{b/m}$                                            & BS/MT beam number                                            \\
	\rowcolor[HTML]{FFFFFF}
	$\theta_{b/m}$                                       & The beamwidth at BS/MT                                       \\
	\rowcolor[HTML]{FFFFFF}
	$\mu_{g}$                                            & The density of beam switch points                            \\
	\rowcolor[HTML]{FFFFFF}
	$K$                                                  & Molecular absorption coefficient                             \\
	\rowcolor[HTML]{FFFFFF}
	$P_{\rm{C}}$                                         & Transmit signal power                                        \\
	\rowcolor[HTML]{FFFFFF}
	$P_{\rm{N}}^{\rm{T}},P_{\rm{N}}^{\rm{M}},P_{\rm{N}}$ & The thermal, molecular absorption and total noise            \\
	\rowcolor[HTML]{FFFFFF}
	{\color[HTML]{000000} $f_c$}                         & {\color[HTML]{000000} Central frequency}                     \\
	\rowcolor[HTML]{FFFFFF}
	$f_{\rm{scs}}$                                       & Subcarrier spacing                                           \\
	\rowcolor[HTML]{FFFFFF}
	$T_{\rm{sym}}$                                       & Symbol length                                                \\
	\rowcolor[HTML]{FFFFFF}
	$\tau$                                               & SSB period                                                   \\
	\rowcolor[HTML]{FFFFFF}
	$U,V$                                                & Frequency, time-domain spacing                               \\
	\rowcolor[HTML]{FFFFFF}
	$N_{\rm{RS}},N_{\rm{SSB}}$                           & Number of REs for RSs and SSBs                               \\
	\rowcolor[HTML]{FFFFFF}
	$N_{s},N_{f}$                                        & Number of REs in time and frequency domain                   \\
	\rowcolor[HTML]{FFFFFF}
	$\alpha$                                             & Time-frequency allocation ratio                              \\
	\rowcolor[HTML]{FFFFFF}
	$B_{\rm{S}},T_{\rm{S}}$                              & The RS bandwidth and time duration                           \\
	\rowcolor[HTML]{FFFFFF}
	$B_{\rm{SSB}},T_{\rm{SSB}}$                          & The SSB bandwidth and time duration                          \\
	\rowcolor[HTML]{FFFFFF}
	$\Delta r,\Delta d_b,\Delta v$                       & The distance, motion and velocity resolutions                \\
	\rowcolor[HTML]{FFFFFF}
	$d_{\rm{max}},v_{\rm{max}}$                          & The unambiguous range and velocity                           \\
	\rowcolor[HTML]{FFFFFF}
	$r_1,r_2,r_i$                                        & Distance to the closest, 2$^{\rm{nd}}\!$ close and interfering BSs \\
	\rowcolor[HTML]{FFFFFF}
	$p_{\rm{m}}^{\rm{s}}$                                & Beam misalignment probability                                \\
	\rowcolor[HTML]{FFFFFF}
	
    \rowcolor[HTML]{FFFFFF}
	$p_{\rm{B}}^{r_1},p_{\rm{B}}^{r_2}$                                    & Blocked probability of the closet and 2$^{\rm{nd}}$ close BSs                                   \\
    \rowcolor[HTML]{FFFFFF}
	$p_{\rm{B}}^{to}$                                    & Timeout probability                                    \\	
    \rowcolor[HTML]{FFFFFF}
    $p_{\rm{I}}$                                         & The BS interfere probability                                 \\	
   \rowcolor[HTML]{FFFFFF}
    $P_{\rm{UB}}^{I}$                                    &  Unblocked probability of the interferer               \\
    \rowcolor[HTML]{FFFFFF}
	$\mathcal{T}$                                        & SINR threshold                                               \\
	$p_{\rm{cvp}}$                                       & The coverage probability                                     \\
\hline
\end{tabular}
\label{table.definition}
\end{table}
\subsection{Network Model}

\begin{figure}[tb]
	\centering
	\includegraphics[width=\linewidth]{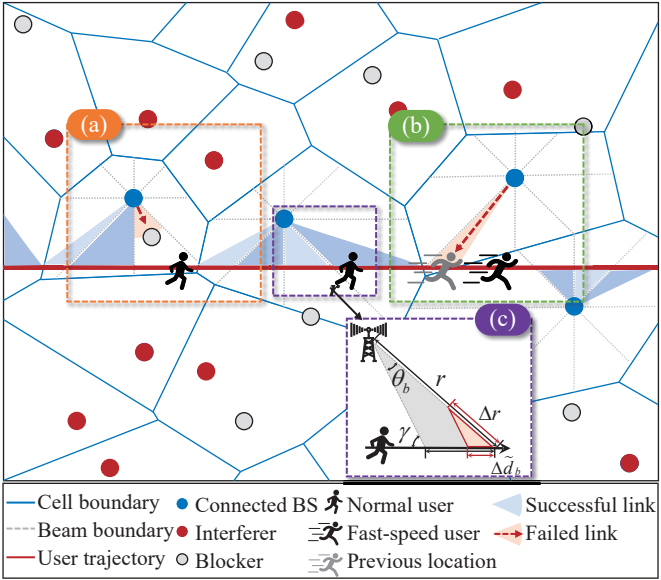}
	\caption{Network deployment, (a) blockage-induced, and (b) movement-induced beam misalignment, and (c) user trajectory.}
	\label{deployment}
\end{figure}

We focus on the outdoor applications of the ISAC-THz networks, considering the impact of the macro-scale movement and LoS blockage.
Based on the well-developed stochastic geometry, the BSs, MTs and blockers are independently distributed following Poisson point process (PPP) distribution with intensity $ \lambda_{\rm{B}},\lambda_{\rm{M}},\lambda_{\rm{S}} $\cite{baccelli2010stochastic}.
The nodes are considered as round dots with the same radius $ r_{\rm{B}} $.
Noticing the LoS-blockage effect at the THz band, the BSs and MTs are also considered as potential blockers.
Directional antennas are used at both the BSs and MTs to enhance the signal power.
The beam number at the BS and MT sides are $ n_b$ and $n_m $, respectively.
We assume the ideal cone beam pattern for simplicity, and the beamwidth at the BS and MT sides are $ \theta_{b}=2\pi/(n_{b})$ and $\theta_{m}=2\pi/(n_{m}) $, respectively\cite{petrov2017interference}.
Considering the widely used maximum receiver power (MRP) scheme, the MT prefers to connect to the closest available BS, forming a Poisson-Voronoi cell tessellation as plotted in Fig.\ref{deployment}\cite{baccelli2009stochastic}.
Without loss of generality, we consider a typical MT moving along a randomly oriented straight line with a speed of $ v $, drawn by the red line.
The beam switch points are the intersections of the beam boundary and user trajectory.

For the described network, the received signal-to-interference-plus-noise-ratio (SINR) at distance $r$ is defined as
\begin{equation}
	{\rm{SINR}}(r)=\frac{P_{\rm{C}}(r)}{I+P_{\rm{N}}},
\end{equation}
where ${P_{\rm{C}}}(r)$ is the received signal power, $ I $ is the interference, and $P_{\rm{N}}$ is the network noise power.
Specifically, we adopt a commonly used THz propagation model
\cite{jornet2011channel}.
The received power from BS$_i$ with a distance of $r_i$ is written as
\begin{align}\label{Pr}
	{P_{\rm{C}}}(r_i) &= {P_{\rm{T}}}{G_b}{G_m}\frac{{{c^2}}}{{{{\left( {4\pi {f_c}} \right)}^2}}}{r_i^{ - 2}}e^{ - K(f_c)r_i} \notag \\
                    &=  A {r_i^{ - 2}}e^{ - K(f_c)r_i},
\end{align}
where $  A \overset{\underset{\mathrm{def}}{}}{=} P_{\rm{T}}G_{b}G_{m}\tfrac{c^{2}}{16\pi^{2}f_c^{2}}$ for simplicity.
$P_{\rm{T}}$ is the transmit power, $ c $ is the speed of light, $ K(f_c) $ is the absorption coefficient depending on the transmission frequency $ f_c $,
and $G_b$ and $G_m$ are the main lobe antenna gains at the BS and MT sides, respectively.
Assuming ideal cone antenna model, the antenna gains $G_b,G_m$ are given as\cite{petrov2017interference}
\begin{equation}
	G_{b/m}=\frac{2}{1-\cos\left(\frac{1}{2}\theta_{b/m}\right)}.
\end{equation}

Due to the blockage effect and the use of directional antennas, not all the BSs will cause interference.
Suppose the interfere probability of BS$_i$ at a distance of $r_i$ is ${p_{\rm{I}}}({r_i})$.
The sum of interference at the typical user can be calculated as
\begin{equation}
	I = \sum\nolimits_{i = 2}^{{N_{{\rm{BS}}}}} {{p_{\rm{I}}}({r_i}) \cdot {P_{\rm{C}}}({r_i})},
	\label{i1}
\end{equation}
where $ N_{\rm{BS}} $ is the number of BSs.


Different from low-band networks, the THz noise consists of the thermal noise $ P_{\rm{N}}^{\rm{T}}$, and the significant molecular absorption noise $ P_{\rm{N}}^{\rm{M}} $\cite{jornet2011channel}.
Induced by the re-radiation of absorbed signal energy, molecular absorption noise depends on the transmit signal power $  {P_{\rm{T}}} $, the number of BS $ N_{\rm{BS}} $, and the beamwidth at both sides $\theta_b, \theta_m$\cite{kokkoniemi2016discussion}.
Using the THz molecular absorption noise model proposed in \cite{kokkoniemi2016discussion}, the network noise power $P_{\rm{N}}$ is
\begin{equation}
	P_{\rm{N}}  =P_{\rm{N}}^{\rm{T}}+P_{\rm{N}}^{\rm{M}},\label{pn_model}
\end{equation}
where $P_{\rm{N}}^{\rm{M}}=\sum\nolimits_{i = 1}^{{N_{{\rm{BS}}}}} \frac{1}{{{n_b}{n_m}}}AK(f_c)r_i^{ - 2}{e^{ - K(f_c){r_i}}}$.

\subsection{Coverage Probability}\label{subsec_covdef}
Coverage probability is defined to characterize the coverage ability and the link quality of the network.
It is the probability that the received SINR exceeds the demodulation threshold $ \mathcal{T} $ under the premise of a successful link connection\cite{baccelli2010stochastic}.
Signal reception requires matched beam pair establishment, called \textit{beam alignment}\cite{kalamkar2021beam}.
Considering the impact of misaligned beams, the coverage probability $ p_{\rm{cvp}}(r_1) $ when connected to the BS$_1$ at a distance of $r_1$ is defined as\cite{chen2021coverage}
\begin{align}
	{p_{{\rm{cvp}}}}(r_1)
	&\!=\! (1 - p_{\rm{m}}^{\rm{s}}) {\mathcal{P}}\left\{ { {{\rm{SINR}}(r_1) > \mathcal{T}} } \right\}\notag\\
	&\!=\! (1 - p_{\rm{m}}^{\rm{s}}) {\mathcal{P}}\left\{ {I+ {P_{\rm{N}}}< \frac{{{P_{\rm{C}}}(r_1)}}{\mathcal{T}}} \!\right\}\!,
	\label{pcvp_concept1}
\end{align}
where $\mathcal{P}(\mathcal{I})$ is the probability of $\mathcal{I}$ is true, and $p_{\rm{m}}^{\rm{s}}$ is the beam misalignment probability.

Eq. (\ref{pcvp_concept1}) shows that the coverage can be enhanced by two means:
1) reducing the beam misalignment probability $p_{\rm{m}}^{\rm{s}}$ by efficient beam management,
and 2) reducing the interference $I$ by suppressing the leak from other BSs.
Thanks to the LoS-dominate THz propagation, the perfectly aligned BSs cause little interference to other nodes.
Therefore, efficient beam alignment improves the THz network coverage in both aspects.

\subsection{Beam Misalignment Probability}\label{subsec_misdef}

Recall that the key task of \textit{beam alignment} is the establishment of suitable beam pair before data transmission\cite{dahlman20205g}.
To that end, 5G new radio (NR) specifies a \textit{measurement-based beam alignment procedure} that consists of three steps, beam-failure detection, candidate-beam identification and recovery-request transmission, summarised as the \textit{detect-error-and-correct method}\cite{dahlman20205g}.
The identification of a new candidate beam relies on the measurement of the periodical SSBs, each of which corresponds to a specific beam and indicates its connectivity\cite{kalamkar2021beam}.
Next, the candidate beam-pair is determined at the BS side based on the SSB measurement reported by the user.
In this way, the user needs to wait for the periodical SSBs to perform measurement and help BS
initiating new candidate-beam selections, which indicates that missed reception of SSBs and untimely beam switch will cause beam misalignment\cite{kalamkar2021beam}.
The main causes include: 1) \textit{LoS blockage} on the BS-MT link forming a blind zone that blocks the signal transmission, as in Fig.\ref{deployment}(a);
2) \textit{User movement} that exceeds the beam boundary, as in Fig.\ref{deployment}(b).
The narrowed THz beamwidth shortens the time it stays within the beam coverage, increasing the probability of beam misalignment and beam switches.

Introducing ISAC into the network, the beam alignment challenge can be alleviated by ISAC-enabled blockage detection and user movement prediction, and thus prevent beam failures by initiating the beam recovery procedure in advance.
Early interventions allow seamless beam switches before beam failure happens, referred to as the \textit{predict-and-prevent method}.
However, the performance of this ISAC approach is limited by the range/velocity sensing resolutions.
Two major causes of beam misalignment in the ISAC-aided network are summarized as follows.
\subsubsection{Imperfect sensing with a probability of $ p_{\rm{err}} $}
Insufficient resolution may cause an estimation error that misleads the BS, resulting in untimely switches and inaccurate beam alignment.
\subsubsection{Association timeout with a probability of $ p_{\rm{B}}^{\rm{to}} $}
Frequent beam switches induce a large time delay that is intolerable for latency-sensitive services, referred to as \textit{association timeout}.

Therefore, for the described ISAC-enabled THz network, the beam misalignment probability $p_{\rm{m}}^{\rm{s}}$ is calculated as
\begin{equation}
	p_{\rm{m}}^{\rm{s}} = p_{\rm{err}}+ p_{\rm{B}}^{{\rm{to}}}.\label{pms_def1}
\end{equation}

Eq.(\ref{pms_def1}) shows that to reduce beam misalignment, a sensing-aided scheme customized for ISAC-enabled THz network and a signal pattern design to reduce the impact of estimation error are of great importance, which are provided in Sec.\ref{sec_JSRS} and Sec.{\ref{sec_optimal_pattern}} respectively.

\section{Joint SSB and RS-Based Sensing-Aided Beam Alignment Scheme}\label{sec_JSRS}
\begin{figure*}[htbp]
	\centering
	\includegraphics[width=\textwidth]{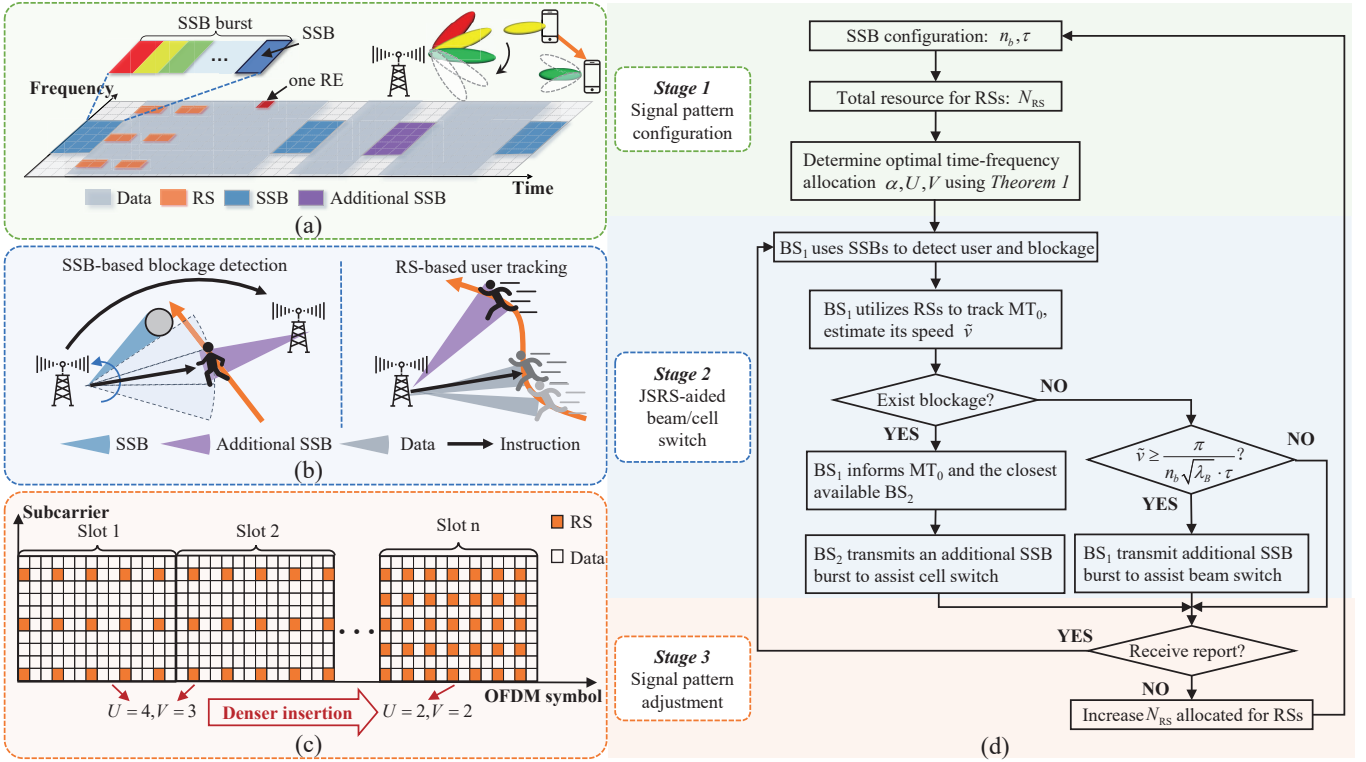}
	\caption{(a) ISAC-frame structure, (b) SSB-based blockage detection and RS-based user tracking, (c) sensing signal configuration and (d) JSRS beam alignment scheme.}
	\label{JSRS}
\end{figure*}
In this section, 
we propose a joint SSB and RS-based sensing scheme to assist THz beam alignment.
Different from the traditional 5G NR specified detect-error-and-correct method, the proposed joint SSB and RS-based sensing scheme
assists the beam switch by a predict-and-prevent-error procedure, thus reducing the beam misalignment.
\subsection{Frame Structure and Sensing Signal Selection}
For better integration with the current 5G air interface, we consider the 5G frame structure and leverage the standard signals to detect potential blockages and track the users to support THz communications.

The 5G-evolved ISAC-frame structure is plotted in Fig.\ref{JSRS}(a), which is based on the current dominant orthogonal frequency division multiplexing (OFDM) waveform\cite{zhang2021design}.
In the time domain, NR transmissions are divided into slots, which are composed of OFDM symbols.
In the frequency domain, the resource is divided by subcarriers.
Thus, the smallest physical resource defined in 5G NR is the \textit{resource element} (RE) that consists of one subcarrier during one OFDM symbol\cite{dahlman20205g}.
NR specifies a beam-swept and time-multiplexed transmission of SSBs, plotted as the blue blocks.
A set of SSBs representing different beam directions is referred to as \textit{SSB burst}.
The reference signals (RSs) inserted in the data are the predefined signals specifically designed for channel estimation and tracking\cite{wei20225g}, plotted as the orange blocks in Fig.\ref{JSRS}(a).
The RSs are comb-type pilot signals that are inserted in data with spacing $ U, V $ in frequency and time respectively\cite{zhang2020perceptive}.

Considering the beam switches are decided at the BS side, downlink sensing is more suitable for the BSs to extract environmental information, because the downlink signals are known to the BSs that can be used for sensing.
However, standard signals are designed for specific functions.
Their bandwidth, duration, sparsity and period affect the sensing abilities\cite{zhang2020perceptive}.
The distance resolution $\Delta r_{\rm{R}}$ depends on the sensing bandwidth $B_{\rm{S}}$ and frequency spacing $U$, whereas the velocity resolution $\Delta v $ depends on the time duration $T_{\rm{S}}$ and time spacing $V$, given as\cite{braun2014ofdm}
\begin{subequations}
	\begin{align}
	&\Delta r_{\rm{R}}=\frac{c}{2UB_{\rm{S}}},\label{drdef}\\
	&\Delta v=\frac{c}{2f_cVT_{\rm{S}}}.\label{dvdef}
	\end{align}
\end{subequations}
Thus, the time and frequency signal mapping affects the joint distance and velocity estimation.
To depict the resource mapping, we define \textit{time-to-frequency allocation ratio} $ \alpha $, which is the ratio of the time domain REs $N_s$ to the frequency domain REs $N_f$.
Specifically, given a total number of $N_{\rm{RS}}$ REs for sensing, allocating an integer $N_s$ OFDM symbols and an integer $N_f$ subcarriers.
It holds true that $N_{\rm{RS}}=N_sN_f$.
The configuration can be expressed as functions of $ \alpha $, i.e, $N_s={N_{\rm{RS}}}^\alpha$ and $N_f={N_{\rm{RS}}}^{1-\alpha}$.
Denoting the subcarrier spacing as  $f_{\rm{scs}}$ and the OFDM symbol length as $T_{\rm{sym}}$,
the bandwidth and time used for RSs are thus
\begin{subequations}
	\begin{align}
		&B_{\rm{S}}=N_ff_{\rm{scs}}={N_{\rm{RS}}}^{1-\alpha} f_{\rm{scs}},\label{bs1}\\
		&T_{\rm{S}}=N_sT_{\rm{sym}}={N_{\rm{RS}}}^\alpha T_{\rm{sym}}.\label{ts1}
	\end{align}
\end{subequations}
The combination of Eq. (\ref{drdef}) and (\ref{bs1}) shows that a small $\alpha$ means a wider beamwidth and better range resolution,
whereas the combination of Eq. (\ref{dvdef}) and (\ref{ts1}) shows a large $\alpha$ means a longer time duration and finer velocity resolution.

User tracking requires high-accuracy range and velocity resolution, whereas relatively coarse range resolution is sufficient for blockage detection.
Balancing the sensing abilities and cost, we select the SSBs for blockage detection and the RSs for user tracking, named the Joint SSB and RS-based sensing (JSRS) scheme.
The reasons are as follows.
\begin{itemize}
	\item Designed for beam synchronization, the periodical SSBs cover all directions.
	Thus, SSB-based sensing enables the BS to realize omni-directional blockage detection with periodically updated information.
	
	\item The RSs inserted in data are user-specific and continuously transmitted during the data phase\cite{3GPP38.213}.
    The RSs span a long time duration with wide bandwidth and can provide better range/velocity resolution for user tracking.
	Thus, they are suitable for constant user tracking, providing real-time positioning information.
	
	\item As pilot signals, both the SSBs and RSs have a good auto-correlation property that leads to improved sensing performance\cite{wei20225g}.
	Additionally, both of them are time and frequency divided with the data payload which causes less interference to data reception.
\end{itemize}

\subsection{Joint SSB and RS-Based Sensing Scheme}\label{scheme}
Different from the traditional detect-interruption-and-correct method, the JSRS scheme follows a predict-and-prevent procedure that provides early intervention to avoid interruption and maintain connection.
The procedures of JSRS are summarized as follows, as depicted in Fig.\ref{JSRS} (d).

\subsubsection{\textbf{Stage 1, sensing signal configuration} }
As in Fig.\ref{JSRS} (a), it starts with initializing the parameters for sensing signal configuration.
The SSB bandwidth $B_{\rm{SSB}}$, duration $T_{\rm{SSB}}$, and the burst period $ \tau $ are selected from 5G numerology.
The number of SSBs within an SSB burst is equal to the beam number $ n_b $.
Configure $ N_{\rm{RS}} $ REs to the RSs, which are allocated with the time-to-frequency ratio $ \alpha $.
The frequency and time insert spacings of the RSs are $ U$ and $V $, respectively.
Parameters $\alpha,U,V$ are obtained using \textit{Theorem \ref{ratio_opt}} given later in Sec.\ref{sec_optimal_pattern}.

\subsubsection{\textbf{Stage 2, JSRS-aided beam/cell switch} }
As plotted in Fig.\ref{JSRS}(b), assistance is realized through three steps, including regular inspection, blockage avoidance and user tracking.

\textit{Step 1: regular inspection.}
During the beam sweeping phase, BS$_1$ utilizes the scheduled periodical SSBs to detect potential blockages in the surroundings and periodically updates the environment information.
During the data transmission phase, BS$_1$ extracts the sensing information from the RSs to estimate the user speed  $ \tilde{v} $.

\textit{Step 2: blockage avoidance.}
If BS$_1$ detects potential blockages, it informs the closest available BS$_2$ to transmit an additional SSB burst and informs the MT to prepare for inter-cell BS handover to BS$_2$.

\textit{Step 3: user tracking.}
If no blockage exists, BS$_1$ further judges whether the MT can receive the regular SSB burst.
Since by the aforementioned measurement-based beam alignment procedure, a user needs to wait for the periodical SSBs to initiate new candidate-beam selection.
Thus, in order to ensure the reception of SSBs and subsequent beam alignment,
the time duration that MT stayed within the beam coverage should be greater than the SSB period $\tau$.
Otherwise, when it holds true that
\begin{equation}
	\tilde{v}\geq\frac{{\bf{E}}[d_b]}{\tau}\stackrel{(a)}{=}\frac{\pi}{ n_b\sqrt{\lambda_B}\cdot \tau},
\end{equation}
BS$_1$ considers to send an additional SSB burst immediately to initiate beam reselection.
Here in (10), the equation $(a)$ comes from the fact that in Poisson Voronoi tessellation, the single beam coverage $ d_b $ follows an exponential distribution with density $ \mu_g=n_b\sqrt{\lambda_B}/\pi $, and ${\bf{E}}[d_b]=1/\mu_g$\cite{kalamkar2021beam}.

Note that in Step 2 and Step 3, the additional SSB burst can be transmitted immediately using the 5G-specified \textit{mini-slot}, which allows the BS to start immediate transmission during the current slot\cite{dahlman20205g}.

\subsubsection{\textbf{Stage 3, pattern adjustment}}
After \textbf{\textit{Stage 2}}, if the BS$_1$ still loses connection to the MT, i.e., no user report is received, the sensing accuracy of JSRS is considered insufficient.
To improve the sensing accuracy, BS$ _1 $ allocates more REs for the RSs to increase the range and velocity resolutions.
Return to \textit{\textbf{Stage 1}} to reconfigure the SSBs and the RS pattern parameters $(\alpha,U,V)$ using \textit{Theorem \ref{ratio_opt}} given later in Sec.\ref{sec_optimal_pattern}.

\section{JSRS Sensing Ability}\label{sec_ability}
In this section, we analyse the effect of time and frequency resource allocation on the signal-sensing abilities for further efficient signal pattern design.
The sensing abilities of concern are the achievable unambiguous range and resolution for distance and velocity.
They are affected by sensing resource allocation, beamwidth and carrier frequency.

\textit{1) Range Resolution:}
It affects both blockage detection and movement estimation.
The range resolution is determined by the signal bandwidth $ B_{\rm{s}}={N_{{\rm{RS}}}}^{1 - \alpha }{f_{{\rm{scs}}}}$ and frequency domain insert spacing $U$, given as \cite{braun2014ofdm}
\begin{align}
\Delta r_{\rm{R}}=\frac{c}{2UB_{\rm{s}}}=\frac{c}{{2U \cdot {N_{{\rm{RS}}}}^{1 - \alpha }{f_{{\rm{scs}}}}}},\label{drrs}
\end{align}
where $N_{\rm{RS}}$ denotes the total REs for sensing, $\alpha$ denotes the time-to-frequency ratio and $f_{\rm{scs}}$ denotes the subcarrier spacing.

Note that (\ref{drrs}) is the resolution at the longitudinal direction \cite{braun2014ofdm}.
However, user tracking is affected by the transverse motion resolution, which is in the crosswise direction to the beam boundary.
As plotted in Fig.\ref{deployment}(c), let $ \theta_b $ be the beamwidth and $ \gamma $ be the angle between the trajectory and the beam boundary.
Since we assume randomly oriented user movement, $ \gamma $ is considered uniformly distributed within $ (\theta_b,\frac{\pi}{2}] $, and thus
the transverse motion resolution $ \Delta \tilde{d}_b $ is also a random variable dependent with $ \gamma $.
In the sequel, we try to derive the average transverse motion resolution, referred to as $ \Delta {d}_b $, for a given longitudinal direction resolution $\Delta r_{\rm{R}}$.
Specifically, by the geometric relationship, we obtain
\begin{equation}
	\frac{\Delta \tilde{d}_b}{\Delta r_R}=\frac{\sin\theta_b}{\sin\gamma},
	\label{geoddb}
\end{equation}
Clearly, from (\ref{geoddb}) it holds true that
\begin{equation}
{\Delta \tilde{d}_b}={\Delta r_R}\frac{\sin\theta_b}{\sin\gamma} . \label{geoddb2}
\end{equation}
By averaging (\ref{geoddb2}) with respect to $\gamma$, we arrive at
\begin{align}
    \Delta {d}_b & \overset{\underset{\mathrm{def}}{}}{=} {\bf{E}}[\Delta \tilde{d}_b]={\Delta r_R}{\bf{E}}[\frac{\sin\theta_b}{\sin\gamma}] \notag\\
	&= {\Delta r_R}\int_{\theta_b} ^{\frac{\pi }{2}}\! {\frac{{\sin \theta_b }}{{\sin \gamma }}\frac{1}{{\frac{\pi }{2} - \theta_b }}d\gamma } \notag\\
	&={\Delta r_R} \frac{{\sin \theta_b }}{{\pi  - 2\theta_b }}\ln\!  \left( {\frac{{1 \!+ \!\cos \theta_b }}{{1\! -\! \cos \theta_b }}} \right).\label{ddb1}
\end{align}
For the ease of exposition, let
\begin{equation}
	A_{\theta} \overset{\underset{\mathrm{def}}{}}{=} \frac{{\sin \theta_b }}{{\pi  - 2\theta_b }}\ln \left( {\frac{{1 +\cos \theta_b }}{{1-\cos \theta_b }}} \right). \label{ddb2}
\end{equation}
Therefore, the motion resolution can be expressed as
\begin{equation}
	\Delta {d_b} = {A_\theta }\Delta r_{\rm{R}}= {A_\theta }\frac{c}{{U \cdot2 {N_{{\rm{RS}}}}^{1 - \alpha }{f_{{\rm{scs}}}}}},\label{ddb}
\end{equation}

\textit{2) Unambiguous Range:}
According to \cite{braun2014ofdm}, the maximum detectable range is determined by the subcarrier spacing $ f_{\rm{scs}} $ and the frequency-domain spacing $ U $, i.e.,
\begin{equation}
	{d_{{\rm{max}}}} = \frac{c}{{U \cdot2 {f_{{\rm{scs}}}}}}.\label{dmax}
\end{equation}

\textit{3) Velocity Resolution:}
For RS-based user tracking, its velocity resolution $ \Delta v $ relates to the central frequency $ f_c $, the time duration $ T_{\rm{S}}={N_{\rm{RS}}}^\alpha T_{\rm{sym}}$ and time-domain spacing $V$.
$ \Delta v $ is\cite{braun2014ofdm}
\begin{equation}
	\Delta v=\frac{c}{2f_cVT_{\rm{S}}}=\frac{c}{2V\cdot f_c{N_{\rm{RS}}}^\alpha T_{\rm{sym}}}.\label{dv}
\end{equation}

\textit{4) Unambiguous Velocity: }
According to \cite{braun2014ofdm}, the maximum detectable velocity is limited by two constraints.
First, the maximum time spacing $ VT_{\rm{sym}}$ is less than the coherence time, namely
\begin{equation}\label{vmaxcon1}
	VT_{\rm{sym}}\leq \frac{c}{2f_cv_{\rm{max}}}.
\end{equation}
Second, industrially, the minimum frequency spacing $Uf_{\rm{scs}}$ is 10 times the maximum Doppler frequency shift\cite{wild2021joint}, namely
\begin{equation}\label{vmaxcon2}
	Uf_{\rm{scs}}\geq\frac{c}{10\cdot2f_cv_{\rm{max}}}.
\end{equation}
Combining (\ref{vmaxcon1}), (\ref{vmaxcon2}), the achievable $ v_{\rm{max}} $ is subjected to
\begin{equation}
	v_{\rm{max}} =\min\left(\frac{U\cdot c f_{\rm{scs}}}{20f_c},\frac{c}{V\cdot2f_c T_{\rm{sym}}}\right).\label{vmax}
\end{equation}

{\textbf{\textit {Example:}}}
Combining (\ref{ddb}), (\ref{dmax}), (\ref{dv}) and (\ref{vmax}), we give an example to show the
sensing abilities of SSBs and RSs in Table \ref{confi_ability}.
Specifically, in this example, the parameters are selected from \cite{tervo20205g,levanen2021mobile}, which provide an evolved numerology from 5G to support THz-band communications.
According to 5GAA, V2X requires 0.1-1 m positioning resolution, covering the intersection with a range of around 100 m, and supporting an average speed of 70 km/h in urban and 120 km/h in highway\cite{5GAA_C-V2X}.
Comparing Table \ref{confi_ability} and these requirements, one can see that
SSBs can detect blockages in V2X networks; meanwhile, RS-based sensing has the potential for localization, and the velocity range can detect vehicles with different mobility.

\begin{table}[tb]
	\centering
	\caption{The sensing abilities of JSRS scheme with reference to 5G numerology.}
	\newcolumntype{Y}{>{\raggedright\arraybackslash}X}	
\begin{tabular}{clllllll}
	\hline
	\multicolumn{8}{c}{\cellcolor[HTML]{DEDEDE}\textbf{System parameters}}                                                            \\ \hline
	\multicolumn{3}{l}{Available bandwidth}                                            & \multicolumn{1}{l|}{1 GHz}       & \multicolumn{3}{l}{SSB period}                                                                                                                                                       & 20 ms                                                            \\
	\multicolumn{3}{l}{Subcarrier spacing}                                             & \multicolumn{1}{l|}{1.92 MHz}    & \multicolumn{3}{l}{SSB duration}                                                                                                                                                     & 17.84 $\mu$s                                                     \\
	\multicolumn{3}{l}{Symbol length}                                                  & \multicolumn{1}{l|}{4.46 $\mu$s} & \multicolumn{3}{l}{SSB bandwidth}                                                                                                                                                    & 240 $f_{\rm{scs}}$                                              \\ \hline
	\multicolumn{8}{c}{\cellcolor[HTML]{DEDEDE}\textbf{SSB-based blockage detection}}                                                                                                                                                                                                                                                                                             \\ \hline
	\multicolumn{4}{c|}{\cellcolor[HTML]{F3F3F3}$d_{\rm{max}}$ (m)}                                                       & \multicolumn{4}{c}{\cellcolor[HTML]{F3F3F3}$\Delta d_b$ (m)}                                                                                                                                                                                            \\ \hline
	\multicolumn{4}{c|}{78.1}                                                                                            & \multicolumn{4}{c}{0.039}                                                                                                                                                                                                                              \\ \hline
	\multicolumn{8}{c}{\cellcolor[HTML]{DEDEDE}\textbf{RS-based user tracking}}                                                                                                                                                                                                                                                                                                   \\ \hline
	\multicolumn{4}{c|}{\cellcolor[HTML]{F3F3F3}$d_{\rm{max}}$ (m)}                                                       & \multicolumn{2}{c|}{\cellcolor[HTML]{F3F3F3}\begin{tabular}[c]{@{}c@{}}$\Delta d_b$ (m)\\ ($B_S$=0.1 GHz)\end{tabular}}       & \multicolumn{2}{c}{\cellcolor[HTML]{F3F3F3}\begin{tabular}[c]{@{}c@{}}$\Delta d_b$ (m)\\ ($B_S$=0.2 GHz)\end{tabular}}     \\ \hline
	\multicolumn{2}{l|}{U=2}                             & \multicolumn{2}{l|}{39.1}                                     & \multicolumn{2}{l|}{0.090}                                                                                                  & \multicolumn{2}{l}{0.045}                                                                                                \\
	\multicolumn{2}{l|}{U=3}                             & \multicolumn{2}{l|}{26.1}                                     & \multicolumn{2}{l|}{0.060}                                                                                                  & \multicolumn{2}{l}{0.030}                                                                                                \\ \hline
	\multicolumn{4}{c|}{\cellcolor[HTML]{F3F3F3}$v_{\rm{max}}$ (km/h)}                                                    & \multicolumn{2}{c|}{\cellcolor[HTML]{F3F3F3}\begin{tabular}[c]{@{}c@{}}$\Delta v$ (m/s)\\ ($T_{\rm{S}}$=0.5 ms)\end{tabular}} & \multicolumn{2}{c}{\cellcolor[HTML]{F3F3F3}\begin{tabular}[c]{@{}c@{}}$\Delta v$ (m/s)\\ ($T_{\rm{S}}$=1 ms)\end{tabular}} \\ \hline
	\multicolumn{2}{c|}{}                                & \multicolumn{1}{l|}{V=1}    & \multicolumn{1}{l|}{550.3}      & \multicolumn{2}{l|}{1.36}                                                                                                   & \multicolumn{2}{l}{0.68}                                                                                                 \\
	\multicolumn{2}{c|}{\multirow{-2}{*}{$f_c$=0.22 THz}} & \multicolumn{1}{l|}{V=3}    & \multicolumn{1}{l|}{183.5}      & \multicolumn{2}{l|}{0.45}                                                                                                   & \multicolumn{2}{l}{0.23}                                                                                                 \\ \hline
	\multicolumn{2}{c|}{}                                & \multicolumn{1}{l|}{V=1}    & \multicolumn{1}{l|}{121.1}      & \multicolumn{2}{l|}{0.30}                                                                                                   & \multicolumn{2}{l}{0.15}                                                                                                 \\
	\multicolumn{2}{c|}{\multirow{-2}{*}{$f_c$=1 THz}}    & \multicolumn{1}{l|}{V=3}    & \multicolumn{1}{l|}{40.36}      & \multicolumn{2}{l|}{0.10}                                                                                                   & \multicolumn{2}{l}{0.05}                                                                                                 \\ \hline
\end{tabular}
	\label{confi_ability}
\end{table}

\begin{remark}
Looking into the expressions in (\ref{ddb}), (\ref{dmax}), (\ref{dv}) and (\ref{vmax}), we arrive at the following observations.
\begin{itemize}
	\item Time-to-frequency allocation: comparing (\ref{ddb}) with (\ref{dv}), a longer signal duration, i.e., the increased $\alpha$, leads to a finer velocity resolution $\Delta v$.
	Wider signal bandwidth, i.e., the decreased $\alpha$, benefits the motion resolution $\Delta d_b$.
	\item Insert spacing $U,V$ selection: comparing (\ref{ddb}) with (\ref{dmax}), a wider frequency spacing $U$ increases the motion resolution $\Delta d_b$, but limits the motion range $d_{\rm{max}}$.
	Similarly, comparing (\ref{dv}) with (\ref{vmax}), larger time spacing $V$ improves velocity resolution $\Delta v$, but reduces velocity range $v_{\rm{max}}$.
	\item Beamwidth selection: observing (\ref{ddb2}) and (\ref{ddb}), the higher angular resolution of a narrow beam enhances the distance estimation, turning the disadvantages of sharp beams into advantages.
\end{itemize}
Clearly, by jointly designing the time-to-frequency ratio $\alpha$ and insert spacings $U,V$, one can optimize the motion and velocity resolution/range jointly.
In the sequel, we move on to sensing signal pattern design that enhances the sensing assistance in beam alignment.
\end{remark}

\section{Sensing Signal Pattern Design}\label{sec_optimal_pattern}
Assisting beam alignment requires a balanced joint resolution of range and velocity.
Thus, in this section, we provide an optimal sensing signal pattern that determines resource allocation in time and frequency domains to minimize the beam misalignment probability while satisfying the sensing requirements for unambiguous range and velocity.
It is noteworthy to note that, since we configure SSBs using the 5G-specified numerology, our focus in this section primarily revolves around the pattern design for the RSs.

\subsection{Problem Formulation}
As defined in Sec.\ref{subsec_misdef}, the beam misalignment in the ISAC-enabled network is induced by imperfect sensing and association timeout.
\begin{itemize}
	\item {Imperfect sensing with a probability of $ p_{\rm{err}} $}:
The performance of JSRS is subjected to perceptual accuracy, including the range and velocity resolutions.
Table.\ref{confi_ability} indicates that the range resolution is sufficient for blockage detection.
Thus, the imperfect sensing-induced estimation error is mainly caused by the limited velocity resolution.
Due to the underestimated user speed, the BS may be unaware of the need for beam switches, causing failures to provide timely assistance.
    \item {Association timeout with a probability of $ p_{\rm{B}}^{\rm{to}} $}:
Considering the delay-sensitive applications, the \textit{association timeout} is defined as two failed attempts to establish the beam pair between BS and MT.
Because the MT prefers to associate with the closest available BS, timeout happens when both the closest and the second-close BSs are blocked.
\end{itemize}
The beam misalignment probability $p_{\rm{m}}^{\rm{s}}$ equals the sum of  $ p_{\rm{err}} $ and $ p_{\rm{B}}^{\rm{to}} $.
By applying PPP distribution probability, we arrive at the following lemma.
\begin{lemma}\label{lemma}
	Consider a JSRS-aided THz network with a motion resolution $ \Delta d_b$, and a velocity resolution $  \Delta v $.
	When the MT speed is $ v $, the BS density is $\lambda_{\rm{B}}$, and the BS beam number is $ n_b $, the beam misalignment probability is
	\begin{align}
		p_{\rm{m}}^{\rm{s}} 
		&=\left({e^{ - {\mu _g}[(v - \Delta v)\tau  - \Delta {d_b}]}} - {e^{ - {\mu _g}v\tau }}\right)\notag\\
		&\cdot{e^{2r_Bw_1 \!+\! \frac{{{w_1}^2}}{{4{\lambda _B}\pi }}}}\left[ {{e^{ - w_2^2}} \!-\! \frac{w_1}{2\sqrt {{\lambda _B}}}{\rm{erfc}}({w_2})} \right]
		\label{pms}\\
		&+ {(2{\lambda _{\rm{B}}}\pi )^2}\int_{2{r_{\rm{B}}}}^{ + \infty } {{r_1}\left( {1 - 	{e^{ - ({\lambda _{\rm{S}}} + {\lambda _{\rm{M}}})({r_1} - 2{r_B})2{r_B}}}} \right)g({r_1})d{r_1}},\notag
	\end{align}
	where $ \Delta d_b, \Delta v $ are given in (\ref{ddb}), (\ref{dv}), ${\mu _g} = {n_b}\sqrt {{\lambda _B}} /\pi ,{w_1} = ({\lambda _{\rm{S}}} + {\lambda _{\rm{M}}})2{r_{\rm{B}}},{w_2} = 2{r_{\rm{B}}}\sqrt {{\lambda _{\rm{B}}}\pi }  + \frac{{{w_1}}}{{2\sqrt {{\lambda _{\rm{B}}}\pi } }}$, and
	\begin{align}
		&g(r_1)
		=\!\!\int_{{r_1}}^{ + \infty } {\left( {1 - {e^{ - ({\lambda _{\rm{S}}} + {\lambda _{\rm{M}}})({r_2} - 2{r_B})2{r_B}}}} \right){e^{ - {\lambda _{\rm{B}}}\pi {r_2}^2}}{r_2}d{r_2}}.\notag
	\end{align}
\end{lemma}
\begin{IEEEproof}
	The proof is given in Appendix \ref{appendix00}.
\end{IEEEproof}

Looking into (\ref{pms}), one can see that the
beam misalignment probability is affected by the sensing resolution through the second exponential term, i.e.,
$ {\Delta v\tau  + \Delta {d_b}} $. 
In addition, according to (\ref{ddb}), (\ref{dv}), the sensing resolutions depend on the sensing signal pattern, that is,
the time-to-frequency resource allocation ratio $ \alpha $ and the insert spacings $ U,V $ which depict the signal sparsity in the frequency and time domains.
Thus, the set of parameters $ (\alpha, U, V) $ that minimizes $ p_{\rm{m}}^{\rm{s}}$ is equivalent to the one
minimizing $ {\Delta v\tau  + \Delta {d_b}}$, i.e.,
\begin{equation}
	 {\rm{arg}}\!\min\limits_{\alpha ,U,V}\! p_{\rm{m}}^{\rm{s}}\!= {\rm{arg}}\!\min\limits_{\alpha ,U,V}\!( {\Delta v\tau \! +\! \Delta {d_b}})  \label{form prob0} 
\end{equation}

Substituting the resolutions (\ref{ddb}) and (\ref{dv}) into (\ref{form prob0}), we formulate the optimization problem as follows.
\begin{subequations}
	\begin{align}
		\!\!(\alpha, U, V) \!=\!  {\rm{arg}}\!\min\limits_{\alpha ,U,V}\! \frac{{c\tau }}{{2{f_c}V \! {N_{{\rm{RS}}}}^\alpha {T_{{\rm{sym}}}}\!}}\! +\! \frac{{c{A_\theta }}}{{2U \! {N_{{\rm{RS}}}}^{1 \!-\! \alpha }\!{f_{{\rm{scs}}}\!}}},\label{form prob}\\
		{\rm{s}}.{\rm{t}}.{v_{{\rm{max}}}}\le \frac{{c \cdot U{f_{{\rm{scs}}}}}}{{20{f_c}}},
		{v_{{\rm{max}}}}\le\frac{c}{{2{f_c} \cdot V{T_{{\rm{sym}}}}}} ,\label{st1}\\
		{d_{{\rm{max}}}}\le\frac{c}{{2U \cdot {f_{{\rm{scs}}}}}} ,\label{st2}\\
		0<\alpha<1,\label{st3}
	\end{align}
\end{subequations}
where (\ref{st1}) is the unambiguous velocity constraint obtained from (\ref{vmax}), and (\ref{st2}) is the unambiguous range constraint obtained from (\ref{dmax}). 

\subsection{Time-Frequency Resource Allocation}
Solving the optimization problem in (24), we get the optimal RS pattern as given in Theorem \ref{ratio_opt}.
%
%

\begin{theorem}
	\label{ratio_opt}
	
	Consider that the ISAC-THz network is required to detect an area with a radius of $ d_{\rm{max}} $ and track users at the maximum speed of $ v_{\rm{max}} $.
	Given ${N_{\rm{RS}}}$ REs for the RSs, define the time-to-frequency allocation ratio $ \alpha $ as the ratio of the time domain REs $N_s$ to the frequency domain REs $N_f$, and the RSs are comb-type signals inserted with $ U$ spacing in frequency and $V$ spacing in time.
	The optimal insert spacings in frequency and time $ U_{\rm{opt}},V_{\rm{opt}} $ and the time-to-frequency allocation ratio $ \alpha_{\rm{opt}} $
    that minimize the beam misalignment probability are
	\begin{align}
		&U_{\rm{opt}} = \left\lfloor {\frac{c}{{2{f_{{\rm{scs}}}}{d_{{\rm{max}}}}}}} \right\rfloor ,V_{\rm{opt}}= \left\lfloor {\frac{c}{{2{f_c}{T_{\rm{sym}}}{v_{{\rm{max}}}}}}} \right\rfloor,\label{UVopt}\\
		&{\alpha _{{\rm{opt}}}} = \frac{1}{2}\left[ {{{\log  }_{{N_{{\rm{RS}}}}}}\left( {\frac{{{U_{{\rm{opt}}}} \cdot {f_{{\rm{scs}}}}\tau }}{{{V_{{\rm{opt}}}} \cdot {f_c}{T_{{\rm{sym}}}}{A_\theta }}}} \right) + 1} \right],\label{alpha_opt}
	\end{align}
where $f_{\rm{scs}}$ is the subcarrier spacing, $T_{\rm{sym}}$ is the symbol length, $f_c$ is the central frequency, $\tau$ is the SSB period, and $A_{\theta}$ is given in (\ref{ddb2}),
decided by the beamwidth $\theta_b$.
\end{theorem}
\begin{IEEEproof}
The proof is given in Appendix \ref{appendix0}.
\end{IEEEproof}

\begin{remark}
Theorem \ref{ratio_opt} instructs the design of the sensing signal pattern,
and reveals its inter-dependency to the beamwidth $\theta_b$, frame structure parameters $f_{\rm{scs}},T_{\rm{sym}},f_c$ and $\tau$.
\end{remark}

%

\section{Coverage Probability}\label{sec_cov_thr}
After the scheme design and resource allocation, we move on to evaluate the benefit of JSRS on the network coverage probability and provide design insights into BS density deployment and beamwidth selection.
In this section, we analyse the coverage probability considering the effect of beam alignment, received signal power, interference, and noise.
\subsection{Interference and Noise Analysis}\label{subsec_inter}

For LoS-dominant THz networks with narrow beams, the network interference is affected by the beam misalignment and LoS blockage\cite{chen2021mobility}.
Thanks to the highly-directional antennas, it is safe to assume that the signal leakage from BSs that are ideally oriented towards their targets is negligible\cite{petrov2017interference}.
Thus, the BS$_i$ interferes with the typical MT when it satisfies three conditions from the following aspects.

\textit{1) The BS is performing beam searching or the BS is misaligned with its intended user during the data transmission phase, whose
probability is referred to as $p_{\rm{I},1}$.}
One SSB period $\tau$ consists of two phases.
At beam searching phase with $T_{\rm{beam}}=n_bT_{\rm{SSB}} $, the scanning beams from the BS may cause interference.
At the data transmission phase with $T_{\rm{data}}=\tau-n_bT_{\rm{SSB}} $, if BS$_i$ is misaligned, it may transmit at other directions, thus causing interference.
Therefore, we obtain
\begin{equation*}
	p_{\rm{I},1}(p_{\rm{m}}^{\rm{s}})\!=\! {\frac{T_{\rm{beam}}}{\tau } \!+ \! {\frac{T_{\rm{data}}}{\tau }}p_{\rm{m}}^{\rm{s}}}\!=\!{\frac{{{n_b}{T_{{\rm{SSB}}}}}}{\tau } \!+ \!\left(  \!{1\! -\! \frac{{{n_b}{T_{{\rm{SSB}}}}}}{\tau }} \! \right) \!p_{\rm{m}}^{\rm{s}}} .
\end{equation*}

\textit{2) The transmit and receive beams of BS$_i$ and MT are accidentally oriented at each other}, whose
probability is
\begin{equation*}
	p_{\rm{I,2}}= \frac{{{\theta _b}}}{{2\pi }}\frac{{{\theta _m}}}{{2\pi }}.
\end{equation*}

\textit{3) The LoS path from BS$_i$ to the typical MT is not blocked.}
In this case, potential blockers include all the other BSs, MTs, and blockers  with a total density of $ \lambda\!=\!\lambda_B\!+\!\lambda_M\!+\!\lambda_S $.
Geometrically, the MT is blocked when a blocker exists on the LoS link and the MT happens to locate within the fan-shaped shadow area of the blocker.
Based on the PPP characteristics, the unblocked probability can be derived as\cite{chen2021coverage}
\begin{equation*}
	\label{pubi}
	p_{\rm{I,3}}(r_i)=P_{{\rm{UB}}}^{\rm{I}}(r_i)= {e^{ - \lambda (r_i - 2{r_B})2{r_B}}}.
\end{equation*}

Therefore, combining 1) - 3), the interfere probability of BS$_i$ is the product of the three probabilities above, i.e.,
\begin{align}
	\!\!p_{\rm{I}}(r_i,p_{\rm{m}}^{\rm{s}})&=p_{\rm{I},1}(p_{\rm{m}}^{\rm{s}})p_{\rm{I},2}p_{\rm{I,3}}(r_i)\notag\\
	&= \left[ {\frac{{{n_b}{T_{{\rm{SSB}}}}}}{\tau } \!+ \!\left(  \!{1\! -\! \frac{{{n_b}{T_{{\rm{SSB}}}}}}{\tau }} \! \right) \!p_{\rm{m}}^{\rm{s}}} \right]  \!\frac{{{\theta _b}}}{{2\pi }}\frac{{{\theta _m}}}{{2\pi }}P_{{\rm{UB}}}^{\rm{I}}(r_i)\! \notag\\
 &= {w_{\rm{s}}} (p_{\rm{m}}^{\rm{s}})P_{{\rm{UB}}}^{\rm{I}}(r_i), \label{pi}
\end{align}
where $ {w_{\rm{s}}}(p_{\rm{m}}^{\rm{s}})\! \overset{\underset{\mathrm{def}}{}}{=} \!\left[ {\frac{{{n_b}{T_{{\rm{SSB}}}}}}{\tau } \!+ \!\left( {1\! -\! \frac{{{n_b}{T_{{\rm{SSB}}}}}}{\tau }} \right)p_{\rm{m}}^{\rm{s}}} \right] \frac{{{\theta _b}}}{{2\pi }}\frac{{{\theta _m}}}{{2\pi }} \!$ for simplicity.

Substituting (\ref{pi}) into the interference signal model (\ref{i1}),
we obtain
\begin{equation}
	I = \sum\nolimits_{i = 2}^{{N_{{\rm{BS}}}}} { {w_{\rm{s}}(p_{\rm{m}}^{\rm{s}})} P_{{\rm{UB}}}^{\rm{I}}(r_i)  \cdot {P_{\rm{C}}}({r_i})}.
	\label{i2}
\end{equation}
Due to the fact that we use the MRP scheme, the MT prefers to connect to the closest available BS,
which yields that $r_i \ge r_1 $. By applying the PPP properties and averaging the interference in (\ref{i2}) with respect to $r_i$, we get \cite{haenggi2009stochastic}
\begin{align}
	{\bf{E}}[I] &= \int_{{r_1}}^{ + \infty }\!\!{2\pi {\lambda _B}r\! \cdot\! {w_{\rm{s}}(p_{\rm{m}}^{\rm{s}})} P_{{\rm{UB}}}^{\rm{I}}(r)  \cdot {P_{\rm{C}}}({r}) } dr\notag\\
    &= \int_{{r_1}}^{ + \infty }\!\!{2\pi {\lambda _B}r\! {w_{\rm{s}}(p_{\rm{m}}^{\rm{s}})} {e^{ - \lambda ({r} - 2{r_B})2{r_B}}} A{r^{ - 2}}{e^{ - K(f_c)r}}} dr\notag\\
	&= 2\pi {\lambda _B}{w_{\rm{s}}(p_{\rm{m}}^{\rm{s}})}{e^{\lambda 4r_B^2}}A \cdot {E_1}[2\lambda {r_B}{r_1} + K(f_c){r_1}],
	\label{eis}
\end{align}
where $ E_{1}[x]\overset{\underset{\mathrm{def}}{}}{=}  \mathop\smallint\nolimits_{x}^{+\infty}r^{-1}e^{-r}dr $ is the exponential integral function.

Next, we analyse the noise in THz networks.
The noise is composed of the constant thermal noise $P_{\rm{N}}^{\rm{T}}$ from the electron devices and the random molecular absorption noise $P_{\rm{N}}^{\rm{M}}$ from the re-emission of the transmit signal power.
Recalling the definition given in (\ref{pn_model}) in Sec.\ref{sec_system}, the noise $P_{\rm{N}}$ is modelled as
\begin{equation}
	\begin{split}
		P_{\rm{N}}&=P_{\rm{N}}^{\rm{T}}+P_{\rm{N}}^{\rm{M}}\\
		&=P_{\rm{N}}^{\rm{T}}+ \sum\nolimits_{i = 1}^{{N_{{\rm{BS}}}}} \frac{K(f_c)}{{{n_b}{n_m}}}Ar_i^{ - 2}{e^{ - K(f_c){r_i}}}.
	\end{split}
\label{pn}
\end{equation}
Similarly, using the PPP property and averaging noise power with respect to $r_i$, we obtain

\begin{align}
	{\bf{E}}[{P_{\rm{N}}}]
	&\!=\! P_{\rm{N}}^{\rm{T}} \!+\! \frac{1}{{{n_b}{n_m}}}\!\!\mathop \int \nolimits_{r_1}^{ \!+ \infty \!}\! \!2\pi {\lambda _B} \cdot AK(f_c){r^{ - 1}}{e^{ \!- K(f_c)r}}dr \notag \\
	&\!=\! P_{\rm{N}}^{\rm{T}} + \frac{{2\pi {\lambda _B}AK(f_c)}}{{{n_b}{n_m}}}{E_1}[K(f_c)r_1]. \label{en}
\end{align}

\subsection{Coverage Probability Analysis}\label{subsec_cov}
Coverage of the MT is deemed achieved when both of the following two events occur:

1) Event $\mathcal{I}_1$: the beams at the BS and MT sides are aligned, the probability of which is $ {\mathcal{P}}(\mathcal{I}_2)=1 - p_{\rm{m}}^{\rm{s}}$.

2) Event $\mathcal{I}_2$:the received SINR exceeds the required demodulation threshold $ \mathcal{T} $, the probability of which is  $ {\mathcal{P}}(\mathcal{I}_1)={\mathcal{P}}\left\{ {\left. {{\rm{SINR}}(r_1) > \mathcal{T}} \right|r_1} \right\}$, where $\!r_1\!$ is the BS$_1$-MT link distance.

Thus, the coverage probability can be expressed as
\begin{align}
	{p_{{\rm{cvp}}}}(r_1)
	&\!=\!{\mathcal{P}}(\mathcal{I}_1) {\mathcal{P}}(\mathcal{I}_2)\notag\\
	&\!=\!(1 - p_{\rm{m}}^{\rm{s}})\cdot {\mathcal{P}}\left\{ {I + P_{\rm{N}}^{\rm{T}}+P_{\rm{N}}^{\rm{M}}< \frac{{{P_{\rm{C}}}(r_1)}}{\mathcal{T}}} \right\}.
	\label{pcvp_concept}
\end{align}
In the expression of (\ref{pcvp_concept}), the thermal noise $P_{\rm{N}}^{\rm{T}}$ is a constant, while the interference $I$ and the molecular absorption noise $P_{\rm{N}}^{\rm{M}}$
are random variables (RVs) with respect to $r_i$, for $i=2,3,\cdots,N_{\rm{BS}}$.

In the sequel, we first separate the random part and the constant part in (\ref{pcvp_concept}), followed by
deriving the cumulative distribution function (CDF) of the random part, thus giving a way to calculate the
coverage probability.

\subsubsection{\textbf{Step 1 Separate variables}}
Given the BS$_1$-MT link distance $r_1$, the molecular absorption noise from the associated BS$_{1}$ can be seen as a constant.
Along with the thermal noise power ${P_{\rm{N}}^{\rm{T}}}$, we define the effective noise $ P_{\rm{N}}^{\rm{eff}}$ as
\begin{align}
	{P_{\rm{N}}^{\rm{eff}}} \overset{\underset{\mathrm{def}}{}}{=} {P_{\rm{N}}^{\rm{T}}} +\frac{K(f_c)}{n_bn_m}\cdot Ar_1^{-2}{e^{-K(f_c){r_1}}}.\label{Con}
\end{align}
Obviously, ${P_{\rm{N}}^{\rm{eff}}}$ is independent with the RVs $r_i, i=2,3,\cdots,N_{\rm{BS}}$.

On the other hand, let $ I_{\rm{eff}}(r_1) $  be the sum of
the interference and the molecular absorption noise from other BSs. It can be written as
\begin{align}
	{I_{\rm{eff}}}=\!\!\sum\limits_{i = 2}^{N_{\rm{BS}} } \left[\left(p_{\rm{I}}(r_i)+\frac{K(f_c)}{n_bn_m}\right) Ar_i^{ - 2}{e^{ -K(f_c){r_i}}}\! \right]\!.\!\label{ieff}
\end{align}

\subsubsection{\textbf{Step 2 Convert expression}} Combined with (\ref{Con}) and (\ref{ieff}), we can re-write
Eq.(\ref{pcvp_concept}) as
\begin{equation}
	{p_{{\rm{cvp}}}}(r_1)\! = \!(1 - p_{\rm{m}}^{\rm{s}}(\alpha,U,V))\mathcal{P}\left\{  I_{{\rm{eff}}}\! < \!\frac{\!{{P_{\rm{C}}}(r_1)}\!}{\mathcal{T}} \!-\! P_{\rm{N}}^{{\rm{eff}}}\! \right\}.\label{pcvp_cdf}
\end{equation}

\subsubsection{\textbf{Step 3 Calculate distribution}}
Based on the PPP properties, the CDF of $ {I_{\rm{eff}}} $ can be obtained using its Laplace function\cite{baccelli2010stochastic}.


With Theorem 1, we have the optimal sensing signal configuration.
Substituting it into (\ref{pcvp_cdf}) and averaging (\ref{pcvp_cdf}) with respect to $ {I_{\rm{eff}}} $, we then get a semi-closed form expression of the coverage probability, as shown in Theorem \ref{theo_pcvp}.

\begin{theorem}\label{theo_pcvp}
	Under the depicted JSRS assisted THz network with a given SINR threshold $ \mathcal{T} $, using the optimal sensing signal configuration $(\alpha,U,V)$ obtained from Theorem \ref{ratio_opt}, the coverage probability of a typical user at the distance of $ r_1 $ from its associated BS is
	\begin{align}
		{p_{{\rm{cvp}}}}({r_1}) &= \!(1
		\!-\! p_{\rm{m}}^{\rm{s}}(\alpha,U,V))\!\int_0^{ + \infty }\! {\frac{{{e^{ - 2\pi {\lambda _B}{f_{\rm{r}}}(s)}}}}{{\pi s}}\!\left[ { - \sin \left( {{f_1}(s)} \right)} \right.} \notag\\
		&\left. { + \sin \left( {2\pi s \cdot \frac{{Ar_1^{-2}{e^{-K(f_c){r_1}}}}}{\mathcal{T}} - {f_1}(s)} \right)} \right]ds,
		\label{pcvp_cal_theo}
	\end{align}
	where $ {f_1}(s) \!=\! -2\pi\!\lambda_B{f_{\rm{i}}}(s) \!-\! 2\pi
	sP_{\rm{N}}^{{\rm{eff}}}$, $ f_{\rm{r}}(s),f_{\rm{i}}(s)$ are
	\begin{subequations}
		\begin{equation}
		\begin{split}
		f_{\rm{r}}(s)  &= \mathop \int \nolimits_{2{r_B}}^{ + \infty } r\left[ {1 - \cos (2\pi sI_{\rm{e}}^{\rm{I}}(r)) \cdot {p_{\rm{I}}}(r,p_{\rm{m}}^{\rm{s}})} \right.\\
		&\left. { - \cos (2\pi sI_{\rm{e}}^{{\rm{UI}}}(r)) \cdot (1 - {p_{\rm{I}}}(r,p_{\rm{m}}^{\rm{s}}))} \right]dr,
		\end{split}
		\label{frs}
		\end{equation}
		\begin{equation}
		\begin{split}
		f_{\rm{i}}(s) &= \mathop \int \nolimits_{2{r_B}}^{ + \infty } r\left[ 	{\sin (2\pi sI_{\rm{e}}^{\rm{I}}(r)) \cdot {p_{\rm{I}}}(r,p_{\rm{m}}^{\rm{s}})} \right.\\
		&\left. { + \sin (2\pi sI_{\rm{e}}^{{\rm{UI}}}(r)) \cdot (1 - {p_{\rm{I}}}(r,p_{\rm{m}}^{\rm{s}}))} \right]dr,		
		\end{split}
		\label{fis}
		\end{equation}
	\end{subequations}
and $I_{\rm{e}}^{\rm{UI}}(r),I_{\rm{e}}^{\rm{I}}(r)$ are, respectively,
\begin{subequations}
	\begin{align}
		&I_{\rm{e}}^{\rm{UI}}(r)=\frac{1}{n_bn_m}AK(f_c){r}^{-2}e^{-K(f_c)r},\notag\\
		&I_{\rm{e}}^{\rm{I}}(r)=A\left(1+\frac{1}{n_bn_m}K(f_c)\right){r}^{-2}e^{-K(f_c)r}.\notag
	\end{align}
\end{subequations}

\end{theorem}
\begin{IEEEproof}
The proof is given in Appendix \ref{appendixC}.
\end{IEEEproof}
\begin{remark}
Theorem \ref{theo_pcvp} can be used to evaluate the coverage enhanced by using JSRS with the optimal sensing signal configuration obtained with Theorem \ref{ratio_opt}.
It reveals that proposed JSRS scheme and improves the ISAC-THz network coverage in two aspects.
From a link perspective, it enhances link quality by reducing beam misalignment probability $p_{\rm{m}}^{\rm{s}}(\alpha,U,V)$ (see Eq.(\ref{pcvp_cal_theo})).
From the network perspective, it reduces the interference and thus increases the SINR, because a better beam alignment helps reduce the interfere probability $p_{\rm{I}}(r,p_{\rm{m}}^{\rm{s}})$ (see Eq(\ref{frs}), (\ref{fis})).

\end{remark}

\section{Numerical Results}\label{sec_numerical}
This section provides numerical results to validate the effectiveness of the JSRS scheme on beam alignment and coverage enhancement.
We also glean design insights in sensing signal configuration, BS density deployment, and beamwidth selection for various conditions.

System parameters are evolved from the 5G numerology to support THz communications, suggested by\cite{tervo20205g,levanen2021mobile}.
Network deployment setups are based on the 5GAA required general V2X use cases, i.e., vehicle communications with the roadside units in urban scenarios\cite{5GAA_C-V2X}.
THz molecular absorption coefficient $K(f_c)$ is obtained from the HITRAN database\cite{gordon2022hitran2020}.
The parameter settings are set as listed in Table \ref{table.parameter}.
Unless otherwise stated, we set the central frequency $f_c=0.34$ THz, the beam number $n_b=128$, the resources for RSs $N_{\rm{RS}}=5e3$,
the BS density $\lambda_{\rm{B}}=2e3/{\rm{km}}^2$, and the user density $\lambda_M=5e3/{\rm{km}}^2$.
Note that, the accuracy of the derived expressions for beam misalignment probability, and coverage probability has been verified by comparing with the Monte-Carlo simulations in our previous works\cite{chen2021coverage,chen2021mobility,chen2022enhancing}.

To illustrate the superiority of JSRS, its performance is compared to the following cases, representing the upper bound, baseline and peer level respectively.
\begin{table}[tb]
	\centering
	\caption{Parameter Settings.}
	\newcolumntype{Y}{>{\raggedright\arraybackslash}X}	
	\begin{tabular}{rl|l}
		\hline
		\rowcolor[HTML]{CECECE}
		\multicolumn{2}{l|}{\cellcolor[HTML]{CECECE}\textbf{System parameters}}                        & \textbf{Value}   \\ \hline
		\rowcolor[HTML]{FFFFFF}
		\multicolumn{1}{r|}{\cellcolor[HTML]{FFFFFF}$f_c$}                 & Central frequency         & $0.1-2$ THz       \\
		\rowcolor[HTML]{FFFFFF}
		\multicolumn{1}{r|}{\cellcolor[HTML]{FFFFFF}$f_{\rm{scs}}$}        & Subcarrier spacing        & $1.92$ MHz        \\
		\rowcolor[HTML]{FFFFFF}
		\multicolumn{1}{r|}{\cellcolor[HTML]{FFFFFF}$T_{\rm{sym}}$}        & OFDM Symbol length        & $4.46\mu$s       \\
		\rowcolor[HTML]{FFFFFF}
		\multicolumn{1}{r|}{\cellcolor[HTML]{FFFFFF}$B_{\rm{tot}}$}        & Available bandwidth       & $1$ GHz           \\
		\rowcolor[HTML]{FFFFFF}
		\multicolumn{1}{r|}{\cellcolor[HTML]{FFFFFF}$T_{\rm{tot}}$}        & Data duration             & $20$ ms           \\
		\rowcolor[HTML]{FFFFFF}
		\multicolumn{1}{r|}{\cellcolor[HTML]{FFFFFF}$N_{\rm{RS}}$}         & The number of REs for RSs & $10^3-10^5$      \\
		\rowcolor[HTML]{FFFFFF}
		\multicolumn{1}{r|}{\cellcolor[HTML]{FFFFFF}$ P_{\rm{T}} $}        & Transmit power            & $23$ dBm          \\
		\rowcolor[HTML]{FFFFFF}
		\multicolumn{1}{r|}{\cellcolor[HTML]{FFFFFF}$P_{\rm{N}}^{\rm{T}}$} & Thermal noise             & $-174$ dBm/Hz     \\ \hline
		\rowcolor[HTML]{CECECE}
		\multicolumn{2}{l|}{\cellcolor[HTML]{CECECE}\textbf{Network deployment\cite{5GAA_C-V2X}}}                       & \textbf{Value}   \\ \hline
		\rowcolor[HTML]{FFFFFF}
		\multicolumn{1}{r|}{\cellcolor[HTML]{FFFFFF}$ r_B $}               & Node radius               & $0.5$ m           \\
		\rowcolor[HTML]{FFFFFF}
		\multicolumn{1}{r|}{\cellcolor[HTML]{FFFFFF}$n_b/n_m$}             & BS/MT beam number         & $32-512$         \\
		\rowcolor[HTML]{FFFFFF}
		\multicolumn{1}{r|}{\cellcolor[HTML]{FFFFFF}$v$}                   & Average speed             & $70$ km/h         \\
		\rowcolor[HTML]{FFFFFF}
		\multicolumn{1}{r|}{\cellcolor[HTML]{FFFFFF}$ \lambda_B $}         & BS density                & $1e3-5e4$ /km$^2$ \\
		\rowcolor[HTML]{FFFFFF}
		\multicolumn{1}{r|}{\cellcolor[HTML]{FFFFFF}$ \lambda_M $}         & MT density                & $1e3-5e4$ /km$^2$ \\
		\rowcolor[HTML]{FFFFFF}
		\multicolumn{1}{r|}{\cellcolor[HTML]{FFFFFF}$ \lambda_S $}         & Blocker density           & $1.5e4$ /km$^2$   \\ \hline
	\end{tabular}
	\label{table.parameter}
\end{table}

\textit{1) Perfect Sensing Case:}
As the upper bound, it stands for the ideal performance of the proposed sensing-aided scheme with no estimation error, labelled as \textit{Perfect sensing}.

\textit{2) 5G Standard Sensing Case:}
As the benchmark, it represents the performance achieved with 5G-required sensing ability, labelled as \textit{5G}.
Based on the 5GAA, the required sensing resolutions for general V2X use cases are set to $ \Delta d_b=0.3$ m in range and $ \Delta V=1$ m/s in velocity\cite{5GAA_C-V2X}.
Substituting the resolutions into Lemma \ref{lemma} and Theorem \ref{theo_pcvp}, we have the beam misalignment probability and coverage probability in 5G cases.	

{3) SSB-Based Sensing Case:}
For peer comparison, we select the single SSB-sensing scheme from work \cite{chen2022enhancing}, labelled as \textit{SSB}.
In this case, SSBs are used for both blockage detection and user tracking.
The motion and velocity resolutions are
\[\Delta d_b=\frac{A_{\theta}c}{2B_{\rm{SSB}}},\Delta v=\frac{c}{2f_cT_{\rm{SSB}}}.\]

\begin{figure}[t]
	\centering
	\includegraphics[width=\linewidth]{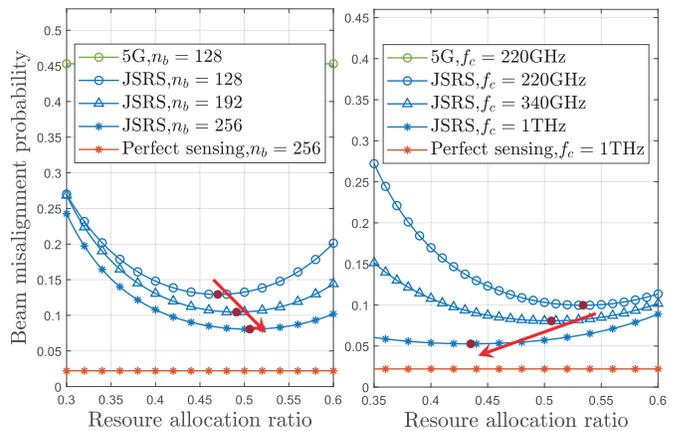}
	\caption{Time-to-frequency allocation ratio affected by (a) central frequency and (b) beamwidth.}
	\label{pm_alpha}	
\end{figure}
\begin{figure*}[tb]
	\centering
	\includegraphics[width=\textwidth]{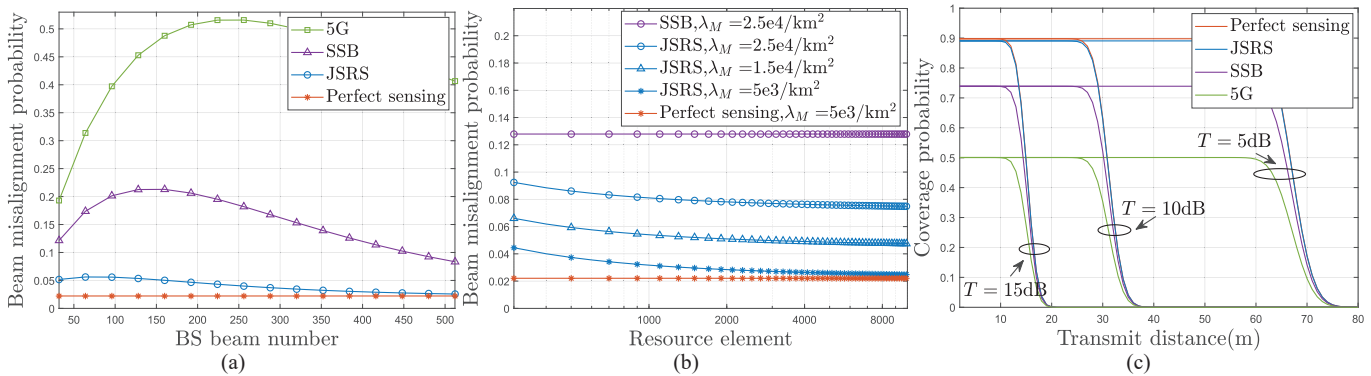}
	\caption{The performance enhancement of the JSRS scheme in the beam alignment and coverage probability.}
	\label{pm_pcvp_cmp}	
\end{figure*}
\subsection{Beam Misalignment Probability}\label{subsec_simbeam}

Averaged in tested cases, JSRS reduces the beam misalignment probability to 0.025, showing an average improvement over 80\% and 70\% to that of 5G and SSB-sensing cases.
We also reveal design insights into the sensing signal pattern design for minimizing the beam misalignment probability.

Fig.\ref{pm_alpha}(a) and Fig.\ref{pm_alpha} (b) reveal design insights in the optimal time-to-frequency allocation ratio to achieve near-ideal performance with fixed sensing resources.
Specifically, Fig.\ref{pm_alpha}(a) shows that utilizing the high angular resolution of narrower beams helps improve the beam alignment and reduce the cost of sensing.
As shown in Fig.\ref{pm_alpha}(a), to cope with the frequent beam switches caused by narrower beams, more resources should be allocated to the time domain for higher velocity resolution, indicated as the increased allocation ratio.
In contrast, Fig.\ref{pm_alpha}(b) shows that more resources should be allocated to bandwidth domain when using higher-frequency signals, thanks to their finer velocity resolution that spares more resources for frequency-domain range perception.

Fig.\ref{pm_pcvp_cmp}(a) shows that the JSRS-scheme significantly reduces the beam misalignment probability, compared to 5G and SSB cases.
The near-ideal performance is achieved when $ n_b\geq 256 $ ($ \theta\leq 1.41^\circ $).
Thanks to the high angular resolution of narrow beams, after reaching the peak, the beam misalignment of the sensing-aided network decreases with the beamwidth.
This suggests that the JSRS scheme turns the disadvantages of sharp beams into advantages, allowing sensing-aided THz networks to use narrower beams to enlarge the coverage.

Fig.\ref{pm_pcvp_cmp}(b) plots the JSRS benefit versus its cost. It shows that JSRS requires a low sensing cost to provide promising assistance,
and outperforms the 5G and SSB-based networks even with insufficient sensing resources.
A near-ideal performance is achieved when $ N_{\rm{RS}}\geq $ 1500, i.e.,
$ T_{\rm{S}}\!=\!$ 0.085 ms, $B_{\rm{S}}\!=\!$ 0.15 GHz.
Such a time-frequency resource setting is acceptable compared to the 10 ms frame length and 1 GHz data bandwidth, where the latter are
defined by current NR standards \cite{3GPP38.213}.
Although a dense-user scenario increases the probability of beam misalignment, it can be compensated by increasing the sensing resource or using narrower beamwidth that both bring in more precise perception.

\begin{figure}[tb]
	\centering
	\includegraphics[width=\linewidth]{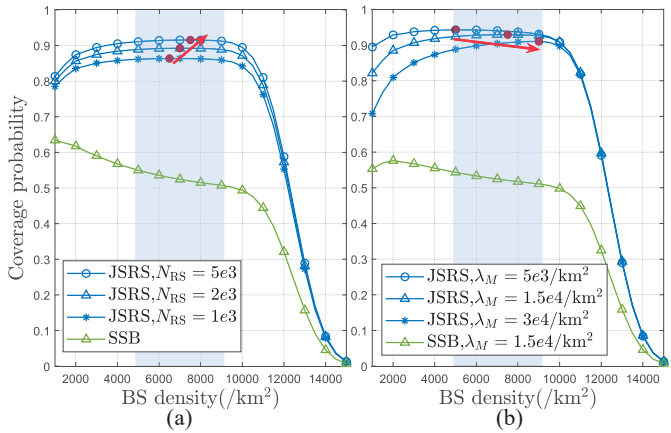}
	\caption{The BS density that maximizes the coverage probability affected by (a) sensing resource and (b) user density. }
	\label{pcvp_lamb}	
\end{figure}
\subsection{Coverage Probability}\label{subsec_simcover}

As shown in Fig.\ref{pm_pcvp_cmp}(c), under tested various scenarios, the JSRS-aided THz network achieves an average coverage probability of 90\%, showing an increment by 75\% and 20\%, when compared to the 5G-case and the SSB-sensing case, respectively.
It achieves near-ideal performance with a small difference in probability of around 0.007.
Satisfying a probability of $ p_{\rm{cvp}}\geq $ 90\%, JSRS-aided THz network covers around 60 m when the SINR threshold is given by $ T\!=\!$ 5 dB, and reaches 30 m when $ T\!=\!$ 10 dB.
Since sensing assistance cannot enhance signal power, the turning points of different cases are similar.
However, the JSRS scheme significantly enhances the coverage by improving the beam alignment, especially within the reachable range of the THz signals.


Fig.\ref{pcvp_lamb} provides design insights in network deployment that maximizes the coverage probability.
In tested cases, preferable BS density lies within $ 5e3-9e3/{\rm{km}} ^2$, marked as the blue strips.
Extra dense BS deployment causes a sharp drop in coverage, e.g., $ \lambda_{B}\geq1e4/{\rm{km}}^2$, suggesting that the THz network needs to restrain the node density to guarantee the link quality.
Specifically, Fig.\ref{pcvp_lamb}(a) shows that the increase of sensing resources allows the network to support denser BSs, because higher sensing accuracy helps mitigate severe LoS-blockage effect on the dense network.
Thus, the JSRS scheme can improve the spatial throughput of THz networks.
As shown in Fig.\ref{pcvp_lamb}(b), the LoS blockage caused by denser users also degrades the coverage probability, requiring more BSs to support massive links.
Increasing the BS density within a reasonable range not only shortens the transmission distance but also reduces the probability of LoS blockage, improving the received signal power with fewer link failures.
Thus, deploying more BSs within a preferable range helps the network support more users.

\section{Conclusion} \label{sec_conclusion}
THz communications can support ultra-high-speed transmission, but it faces a coverage bottleneck that requires sharp beams to compensate for the propagation loss, which in turn brings in the narrow beam alignment challenge.
In line with the 5G beam management, this paper proposed a JSRS scheme to assist THz beam alignment, which is highly compatible with the 5G air interface.
Different from the traditional detect-and-correct method, JSRS helps the BS predict the impending beam switches and thus prevent beam misalignment, enhancing the network coverage probability.
We provided a sensing signal configuration that determines the optimal time-to-frequency pattern to minimize beam misalignment with fixed sensing resources.
Based on stochastic geometry, we derived the coverage probability of the ISAC-THz network, which shed light on the network deployment and can be used to evaluate the performance gain
by integrating sensing.
These insights also play important roles in follow-up studies on sensing resource allocation.
In tested urban V2X scenarios, numerical results demonstrated that JSRS achieves near-ideal performance and effectively reduces average 80\% beam misalignment, and increases 75\% coverage probability, compared to 5G cases.

\begin{appendices}
\section{}\label{appendix00}
Recall the definition of beam misalignment probability.
It is composed by the probabilities of imperfect sensing $p_{\rm{err}}$ and timeout $p_{\rm{B}}^{\rm{to}}$.
\begin{equation}
	p_{\rm{m}}^{\rm{s}} = p_{\rm{err}}+ p_{\rm{B}}^{{\rm{to}}}.\label{pms_def2}
\end{equation}

First, we analyse the timeout probability $p_{\rm{B}}^{\rm{to}}$.
It is defined as two failed attempts to establish the beam pair between BS and MT, i.e., when both BS$_1$ and BS$_2$ are blocked.
Suppose the distance to the closest BS$_1$ and second close BS$_2$ are $r_1$ and $r_2$, respectively.
Using the PPP distribution properties, the joint distance probability density function (pdf) of $r_1, r_2$ is\cite{moltchanov2012distance}
\begin{equation}
	{f_{{r_1},{r_2}}}({r_1},{r_2})
	={e^{ - {\lambda _{\rm{B}}}\pi {r_2}^2}}{(2{\lambda _{\rm{B}}}\pi )^2}{r_1}{r_2},
	\label{fr1r2_joint}
\end{equation}
where $ \lambda _{\rm{B}} $ is the BS density.
Averaging over the joint PDF of $r_1,r_2$, the average timeout probability can be expressed as
\begin{equation}
	p_{\rm{B}}^{{\rm{to}}}= \int_{2{r_{\rm{B}}}}^{ + \infty } {\int_{{r_1}}^{ + \infty } {{p_{\rm{B}}}({r_1}){p_{\rm{B}}}({r_2}){f_{{r_1},{r_2}}}({r_1},{r_2})d{r_2}} d{r_1}},
	\label{pto_def}
\end{equation}
where the inner lower bound is because $r_2>r_1$, the outer lower bound is because BS and MT will not overlap, i.e., $r_1> 2r_{\rm{B}} $, and
${p_{\rm{B}}}({r})$ denotes the LoS blockage probability of a BS located at the distance $r$.
It is the probability that there exist at least one blocker or MT that blocked the BS transmission.
Based on PPP property, it is\cite{chen2021coverage}
\begin{equation}
	{p_{\rm{B}}}({r})={1 - {e^{ - (\lambda _{\rm{S}} + \lambda _{\rm{M}})({r} - 2{r_B})2{r_B}}}},
	\label{pbr}
\end{equation}
where $\lambda _{\rm{S}}$ is the blocker density and $\lambda _{\rm{M}} $ is the MT density.
Substituting (\ref{fr1r2_joint}) and (\ref{pbr}) into (\ref{pto_def}), we have
\begin{equation}
	\begin{split}
		p_{\rm{B}}^{{\rm{to}}} &= \int_{2{r_{\rm{B}}}}^{ + \infty } {\left[ {{{(2{\lambda _{\rm{B}}}\pi )}^2}{r_1}\left( {1 - {e^{ - ({\lambda _{\rm{S}}} + {\lambda _{\rm{M}}})({r_1} - 2{r_B})2{r_B}}}} \right)} \right.} \\
		&\left. { \cdot \int_{{r_1}}^{ + \infty } {\left( {1 - {e^{ - ({\lambda _{\rm{S}}} + {\lambda _{\rm{M}}})({r_2} - 2{r_B})2{r_B}}}} \right){e^{ - {\lambda _{\rm{B}}}\pi {r_2}^2}}{r_2}d{r_2}} } \right]d{r_1}\\
		&= {(2{\lambda _{\rm{B}}}\pi )^2}\int_{2{r_{\rm{B}}}}^{ + \infty } {{r_1}\left( {1 - {e^{ - ({\lambda _{\rm{S}}} + {\lambda _{\rm{M}}})({r_1} - 2{r_B})2{r_B}}}} \right)g({r_1})d{r_1}},
		\label{epto_joint}
	\end{split}
\end{equation}
where $g(r_1)$ is the inner integration
\begin{equation}
	{g(r _1)=\int_{{r_1}}^{ + \infty } {\left( {1 - {e^{ - ({\lambda _{\rm{S}}} + {\lambda _{\rm{M}}})({r_2} - 2{r_B})2{r_B}}}} \right){e^{ - {\lambda _{\rm{B}}}\pi {r_2}^2}}{r_2}d{r_2}} }.
\end{equation}

Next, we analyse the probability $p_{\rm{err}}$ of imperfect sensing-induced beam misalignment.
As shown in Table \ref{confi_ability}, $ \Delta d_b \ll  r_{\rm{B}} $, indicating that the range resolution is sufficient to detect the blockages.
The imperfect sensing-induced beam misalignment happens when the BS$_1$ is not blocked but it underestimates the MT speed that is unaware of the need to send an extra SSB burst to assist beam reselection with a probability of $p_{{\rm{ve}}}^{\rm{s}}$.
Because the distribution of MT speed $v$ and distance $r_1$ are independent, $p_{\rm{err}}$ can be expressed as 
\begin{equation}
	{p_{{\rm{err}}}} = p_{{\rm{ve}}}^{\rm{s}}(1 - {\bf{E}}[p_{\rm{B}}^{{r_1}}]).
	\label{perr_append}
\end{equation}
Considering uniformly distributed $v$, probability $ p_{{\rm{ve}}}^{\rm{s}}$ equal to the case when the actual speed satisfies $ v>d_b/\tau $, but the estimated speed satisfies $ \tilde{v}=v-\Delta v<d_b/\tau $.
The corresponding probability can be expressed as
\begin{align}
	p_{{\rm{ve}}}^{\rm{s}}
	&=\mathcal{P}\left\{ {v > \frac{{{d_b}}}{\tau},v - \Delta v < \frac{{{d_b} + \Delta {d_b}}}{\tau }} \right\}\notag\\
	&\stackrel{(a)}{=} {e^{ - {\mu _g}[(v - \Delta v)\tau  - \Delta {d_b}]}} - {e^{ - {\mu _g}v\tau }},	
	\label{pve}
\end{align}
where $(a)$ results from the fact that the beam boundaries form Poisson Voronoi tessellation and the single beam coverage $ d_b $ follows an exponential distribution with density $ \mu_g=n_b\sqrt{\lambda_B}/\pi $\cite{kalamkar2021beam}.

Because the BSs are PPP distributed, the closest BS distance $r_1$ follows the PDF of ${f_{{r_1}}}({r_1}) = 2\pi {\lambda _B}{r_1}{{\mathop{\rm e}\nolimits} ^{ - {\lambda _B}\pi {r_1}^2}}$\cite{moltchanov2012distance}.
Averaging (\ref{pbr}) over $r_1$, the BS$_1$ blocked probability is
\begin{align}
	{\bf{E}}[p_{{\rm{B}}}^{{r_1}}]
	&=\int_{2r_{\rm{B}}}^{+\infty}	{p_{{\rm{B}}}}(r_1){f_{{r_1}}}(r_1)dr_1\notag\\
	&=1\!-\!{e^{2r_Bw_1 \!+\! \frac{{{w_1}^2}}{{4{\lambda _B}\pi }}}}\left[ {{e^{ - w_2^2}} \!-\! \frac{w_1}{2\sqrt {{\lambda _B}}}{\rm{erfc}}({w_2})} \right],
	\label{ebr1}
\end{align}
where  $w_1\!=\!(\lambda_{S}\!+\!\lambda_{M})2r_{\rm{B}},w_2 \!= \!2{r_{\rm{B}}}\sqrt {{\lambda _B}\pi }\! +\!\!\frac{w_1}{{2\sqrt {{\lambda _B}\pi } }}$, and
$ {\rm{erfc}}(x)=\tfrac{2}{\sqrt{\pi}}\int_{x}^{+\infty}e^{-x^2}dx $ is the complementary error function.		
Substituting (\ref{pve}) and (\ref{ebr1}) into (\ref{perr_append}), we have the imperfect sensing-induced beam misalignment probability
\begin{align}
	p_{{\rm{err}}}=
	&\left({e^{ - {\mu _g}[(v - \Delta v)\tau  - \Delta {d_b}]}} - {e^{ - {\mu _g}v\tau }}\right)\notag\\
	&\cdot{e^{2r_Bw_1 \!+\! \frac{{{w_1}^2}}{{4{\lambda _B}\pi }}}}\left[ {{e^{ - w_2^2}} \!-\! \frac{w_1}{2\sqrt {{\lambda _B}}}{\rm{erfc}}({w_2})} \right].
	\label{perr_derived}
\end{align}

Finally, substituting (\ref{epto_joint}) and (\ref{perr_derived}) into (\ref{pms_def2}), we have the beam misalignment probability given in (\ref{pms}).

\section{}\label{appendix0}
Recalling the function to be minimized in (\ref{form prob}), we first obtain the spacings $U,V$ and then determine the ratio $\alpha$.

Given $N_{\rm{RS}}$, the larger the insert spacing $ U, V $, the smaller the $g(\alpha,U,V)$.
Therefore, the optimal $ U_{\rm{opt}}, V_{\rm{opt}} $ are the largest integers satisfying the constraints for sensing unambiguous range and velocity (see (\ref{st1}), (\ref{st2})), given as
\begin{align}
	U_{\rm{opt}} = \left\lfloor {\frac{c}{{2{f_{{\rm{scs}}}}{d_{{\rm{max}}}}}}} \right\rfloor,V_{\rm{opt}}= \left\lfloor {\frac{c}{{2{f_c}{T_{\rm{sym}}}{v_{{\rm{max}}}}}}} \right\rfloor,\label{uoptvopt}
\end{align}
respectively.
By setting $U=U_{\rm{opt}}$ and $V=V_{\rm{opt}}$, we notice that $ g(U_{\rm{opt}},V_{\rm{opt}},\alpha)$ is a convex function about $ \alpha $, the derivative of which with
respect to $ \alpha $ is
\begin{equation}
	\begin{split}
 	g'(U,V,\alpha )= -\frac{{c\tau {N_{{\rm{RS}}}}^{ - \alpha }\ln ({N_{{\rm{RS}}}})}}{{V2{f_c}{T_{{\rm{sym}}}}}} + \frac{{A_\theta }c{N_{{\rm{RS}}}}^{\alpha - 1}}{{U2{f_{{\rm{scs}}}}}}\ln ({N_{{\rm{RS}}}}).
	\end{split}
\end{equation}
The optimal $ \alpha_{\rm{opt}} $ appears when  $ g'(U_{\rm{opt}},V_{\rm{opt}},\alpha ) =0  $, i.e.,
\begin{align}
	&\frac{{\tau {N_{{\rm{RS}}}}^{ - {\alpha _{{\rm{opt}}}}}}}{{V \cdot 2{f_c}{T_{{\rm{sym}}}}}} = \frac{{{A_\theta }}}{{U \cdot 2{f_{{\rm{scs}}}}}}{N_{{\rm{RS}}}}^{{\alpha _{{\rm{opt}}}} - 1}\notag\\
\Leftrightarrow&{N_{{\rm{RS}}}}^{2{\alpha _{{\rm{opt}}}} - 1} = \frac{{U \cdot \tau {f_{{\rm{scs}}}}}}{{V \cdot
		{f_c}{T_{{\rm{sym}}}}{A_\theta }}}.\label{galpha}
\end{align}
Based on (\ref{galpha}), we derive the expression for the optimal $ \alpha_{\rm{opt}} $
\begin{equation}
	\begin{split}
	{\alpha _{{\rm{opt}}}} = \frac{1}{2}\left[ {{{\log }_{{N_{{\rm{RS}}}}}}\frac{{U_{\rm{opt}}\cdot{f_{{\rm{scs}}}}\tau }}{{V_{\rm{opt}}\cdot{f_c}{T_{{\rm{sym}}}}{A_\theta }}} + 1} \right].
	\end{split}
\end{equation}
\section{}\label{appendixC}
The coverage analysis in this paper is tailored for ISAC-THz networks and shows the gain of using sensing to assist networking.
It makes innovations in the following aspects.
First, we consider the randomness of the THz molecular absorption noise in the modelling.
Second, we model the effect of beam misalignment on the network coverage.
We consider misalignment caused by LoS blockage, imperfect sensing and association timeout.
Third, we model the benefit of sensing-assistance that can be used to evaluate the gain of introducing sensing into THz networks.

We calculate the coverage probability following the 3 steps summarized in Sec.\ref{subsec_cov}, i.e., variable separation, expression conversion and distribution calculation.
\color{black}

Recall the coverage probability $ p_{\rm{cvp}}(r_1) $ written by the CDF of $I_{\rm{eff}}$, as given in (\ref{pcvp_cdf})
\begin{align}
	{p_{{\rm{cvp}}}}(r_1)
	\!=\! (1 - p_{\rm{m}}^{\rm{s}})\mathcal{P}\left\{ {I_{{\rm{eff}}} \!< \!{ \frac{{{P_{\rm{C}}}(r_1)}}{\mathcal{T}}}\! -\! P_{\rm{N}}^{{\rm{eff}}}} \!\right\}\!.\! \label{pcvp_app}
\end{align}
Since $p_{\rm{m}}^{\rm{s}}$ is obtained using Lemma \ref{lemma}, we derive the CDF of $ I_{\rm{eff}} $ over the BS distribution, which can be written as
\begin{equation}
	{p_{{\rm{cm}}}} = \mathcal{P}\left\{ { {I_{{\rm{eff}}} < \frac{{{P_{\rm{C}}}(r_1)}}{\mathcal{T}} - P_{\rm{N}}^{{\rm{eff}}}} } \right\}.\label{pcm_concept}
\end{equation}
Thanks to the homogenous PPP distribution of the BSs, the effective interference $I_{\rm{eff}}$ to a typical user forms a shot-noise field\cite{chen2021coverage}.
Based on this property, $p_{\rm{cm}}$ can be calculated as\cite{baccelli2009stochastic}
\begin{equation}
	\begin{split}
		\!\!{p_{{\rm{cm}}}}\!=\! \!\mathop\int_{\! -\! \infty }^{ \!+\! \infty }\!\! {{{\cal L}_{\rm{I}}}}\! \left( {s,{r_1}} \right){\!e^{ - 2i\pi sP_{\rm{N}}^{{\rm{eff}}}}}\frac{{{e^{\frac{{2i\pi s P_{\rm{C}}({r_1})}}{\mathcal{T}}}} \!-\! 1}}{{2i\pi s}}ds,\!\!
		\label{pcm_theo}
	\end{split}
\end{equation}
where ${{\mathcal{L}}}_{\rm{I}}({s,{r_1}})={\bf{E}}\left[{e^{{ - 2i\pi s{I_{\rm{eff}}}}}}\right]  $ is the Laplace functional of $ I_{\rm{eff}} $.
Using the expression of $ I_{\rm{eff}}$ given in (\ref{ieff}), we have
\begin{equation}
	\begin{split}
	\!\!{{\cal L}_{\rm{I}}}(s,{r_1})
    \!\stackrel{(a)}{=} \!{\bf{E}}\!\left[{e^{ - 2i\pi s\!\sum\! f ({r_i})}}\right]\!\stackrel{(b)}{=}\! \prod\limits_{i = 2}^{{N_{{\rm{BS}}}}}\! {{{\bf{E}}}}\! \left[ {{e^{ - 2i\pi sf({r_i})}}} \right],\label{lieff1}
	\end{split}
\end{equation}
where in (a), we define $f(r_i)$ for simplicity
\begin{equation}
	f({r_i}) = \left( {{a_j} + K} \right)A{r_i}^{ - 2}{e^{ - K{r_i}}},\label{fri}
\end{equation}
and where the $\{a_j\}$ is a set of RVs with binomial distribution $ \mathcal{P}\{a_j=1\}=p_{\rm{I}}(r_i),\mathcal{P}\{a_j=0\}=1-p_{\rm{I}}(r_i) $
(see (\ref{pi}) for $p_{\rm{I}}(r_i)$).
Equation (b) in (\ref{lieff1}) results from the fact that the BSs are independently distributed.
Using the probability generating functional (PGFL) of the PPP\cite{baccelli2009stochastic}, (\ref{lieff1}) is then given by
\begin{align}
	\!{{\mathcal{L}}}_{\rm{I}}({s,{r_1}}\!)\!=\!\!\exp \!\left( {\! -\! 2\pi {\lambda_B}\!\!\mathop \int \nolimits_{r_1}^{ \!+ \!\infty \!\!}\!\! r\!\left(\! {1\! -\! {{\bf{E}}}[{e^{\! -\! 2i\pi sf(r)}}]}\right)\!dr}\!\! \right)\!\!,\!\label{lieffs}
\end{align}
where the appearance of the lower bound $r_i \ge r_1 $ is due to the fact the MT prefers to connect to the closest available BS
and the interferers are farther away from the MT.
Averaging ${e^{ - 2i\pi sf(r)}}$ in (\ref{lieffs}) with respect to $a_j$, we get
\begin{equation}
	\begin{split}
		\!{{\bf{E}}}[{e^{ \!- \!2i\pi s\!f(r)}}] \!= \!{e^{ \!- 2i\pi\! sI_{{\rm{e}\!}}^{\rm{I}}(r)}}
		p_{\rm{I}}(r) \!+ \!{e^{ \!- 2i\pi\! s I_{{\rm{e}}}^{{\rm{UI}}}(r)}}(1\! -\! p_{\rm{I}}(r)),\!\!
	\end{split}\label{eas}
\end{equation}
where  $p_{\rm{I}}(r))$ is given in (\ref{pi}), $ I_{\rm{e}}^{\rm{UI}}(r)$ and $I_{\rm{e}}^{\rm{I}}(r)$ are, respectively, given by
\begin{equation}
I_{\rm{e}}^{\rm{UI}}(r)=\frac{AK}{n_bn_m}{r}^{-2}e^{-Kr},I_{\rm{e}}^{\rm{I}}(r)=\left(A+\frac{AK}{n_bn_m}\right){r}^{-2}e^{-Kr}.\notag
\end{equation}

Next, use the Euler's formula, i.e., $ e^{2i\pi x}=\cos(2\pi x)+i\sin(2\pi x) $\cite{chen2021coverage}, to separate the real and imaginary component in (\ref{eas}), and rewrite it as
\begin{align}
	\!{{\bf{E}}}[{e^{ \!- \!2i\pi s\!f(r)}}] &\!= \!
	\left[\cos (2\pi sI_{\rm{e}}^{\rm{I}}(r))\!+\!i\sin(2\pi sI_{\rm{e}}^{\rm{I}}(r)) \right] {p_I}(r)\label{ea_eular}\\
	&\!+ \!\left[\cos (2\pi sI_{\rm{e}}^{\rm{UI}\!}(r))\!+\! i\!\sin(2\pi sI_{\rm{e}}^{\rm{UI}\!}(r)) \right]\!(1\! -\! p_{\rm{I}}(r)).\!\!	\notag
\end{align}
Substituting (\ref{ea_eular}) into (\ref{lieffs}), we have
\begin{equation}
	{{\cal L}_I}(s,r_1) = \exp \left[-2\pi\lambda_B\left(f_{\rm{r}}(s) + i\cdot f_{\rm{i}}(s)\right)\right] ,\label{lieff_f}
\end{equation}
where $ f_{\rm{r}}(s), f_{\rm{i}}(s) $ are the real and imaginary parts, given by
\begin{subequations}
\begin{align}
	f_{\rm{r}}(s)  =& \mathop \int \nolimits_{2{r_B}}^{ + \infty } r\left[ {1 - \cos (2\pi sI_{\rm{e}}^{\rm{I}}(r)) \cdot {p_I}(r)} \right.\notag\\
	&\left. { - \cos (2\pi sI_{\rm{e}}^{{\rm{UI}}}(r)) \cdot (1 - {p_I}(r))} \right]dr,\label{fcs}\\
	f_{\rm{i}}(s) =& \mathop \int \nolimits_{2{r_B}}^{ + \infty } r\left[ {\sin (2\pi sI_{\rm{e}}^{\rm{I}}(r)) \cdot {p_I}(r)} \right.\notag\\
	&\left. { + \sin (2\pi sI_{\rm{e}}^{{\rm{UI}}}(r)) \cdot (1 - {p_I}(r))} \right]dr,\label{fss}
\end{align}	
\end{subequations}
respectively.
Using the parity of the trigonometric functions,  we find that $ f_{\rm{r}}(s) $ is even and $ f_{\rm{i}}(s) $ is odd.
Thus, using the conjugate property of (\ref{lieffs})), we have
\begin{equation}
	{\cal L}_I( -s,{r_1}) = \exp \left[ { - 2\pi {\lambda _B}\left( {f_{\rm{r}}(s) - i\cdot f_{\rm{i}}(s)}\right)} \right],
	\label{lieff_f_-s}
\end{equation}
Substituting (\ref{lieff_f}) and (\ref{lieff_f_-s}) into (\ref{pcm_theo}), the integrand function  of $ p_{\rm{cm}} $ can be written as
\begin{equation}
	{f_{{\rm{int}}}}(s) = {e^{-2\pi\lambda_B{f_{\rm{r}}}(s)}}\frac{{{e^{i\cdot{f_2}(s)}} - {e^{  i\cdot{f_1}(s)}}}}{{2i\pi s}},\label{fint}
\end{equation}
where $ f_1(s),f_2(s) $ are given as
\begin{subequations}
\begin{align}
	{f_1}(s)& = -2\pi\lambda_B{f_{\rm{i}}}(s) - 2\pi sP_{\rm{N}}^{{\rm{eff}}},\label{f1}\\
	{f_2}(s) &= 2\pi s\cdot\frac{{{P_C}({r_1})}}{\mathcal{T}} -2\pi\lambda_B {f_{\rm{i}}}(s) - 2\pi sP_{\rm{N}}^{{\rm{eff}}},\label{f2}
\end{align}	
\end{subequations}
where $P_{\rm{N}}^{{\rm{eff}}}$ is given in (\ref{Con}).
As $ f_i(s) $ is an odd function, observing equations (\ref{f1}) and (\ref{f2}), it follows that $ f_1(s)$ and $f_2(s) $ are odd functions about $s$.
By noticing that the symmetry of the integral interval and substituting (\ref{fint}) into (\ref{pcm_theo}), we obtain
\begin{equation}	
	\begin{split}
		{p_{{\rm{cm}}}}
		&=\int_{-\infty}^{ + \infty }  {f_{{\rm{int}}}}(s) ds\\
		&= \int_0^{ + \infty }  [{f_{{\rm{int}}}}(s) + {f_{{\rm{int}}}}( - s)]ds\\
		&= \int_0^{ + \infty }  \frac{{{e^{-2\pi\lambda_B{f_{\rm{r}}}(s)}}}}{{\pi s}}\left[ {\sin ({f_2}(s)) - \sin ({f_1}(s))} \right]ds.
	\end{split}\label{pcm_cal}
\end{equation}
Inserting (\ref{pcm_cal}) into (\ref{pcvp_app}) gives
\begin{align}
		\!\!{p_{\rm{cvp}}}(r_1)\!
	&\!= \!(1\! -\! p_{\rm{m}}^{\rm{s}})\cdot{p_{{\rm{cm}}}} \\\label{pcvp_cal}
	&\!=\!(\!1 \!-\! p_{\rm{m}}^{\rm{s}}\!)\! \!\!\mathop\int\nolimits_0^{ \!+ \!\infty\!}\! \frac{{{e^{\!-\!2\pi\!\lambda_B\!f_{\rm{r}}(s)}}\!\!}}{{\pi s}}\left[ {\sin ({f_2}(s)) \!-\! \sin ({f_1}(s))} \right]\!ds.\!\!\notag
\end{align}
\end{appendices}
\bibliography{ISAC_coverage_throughput}
\bibliographystyle{IEEEtran}
\begin{IEEEbiography}[{\includegraphics[width=1in,height=1.25in,keepaspectratio]{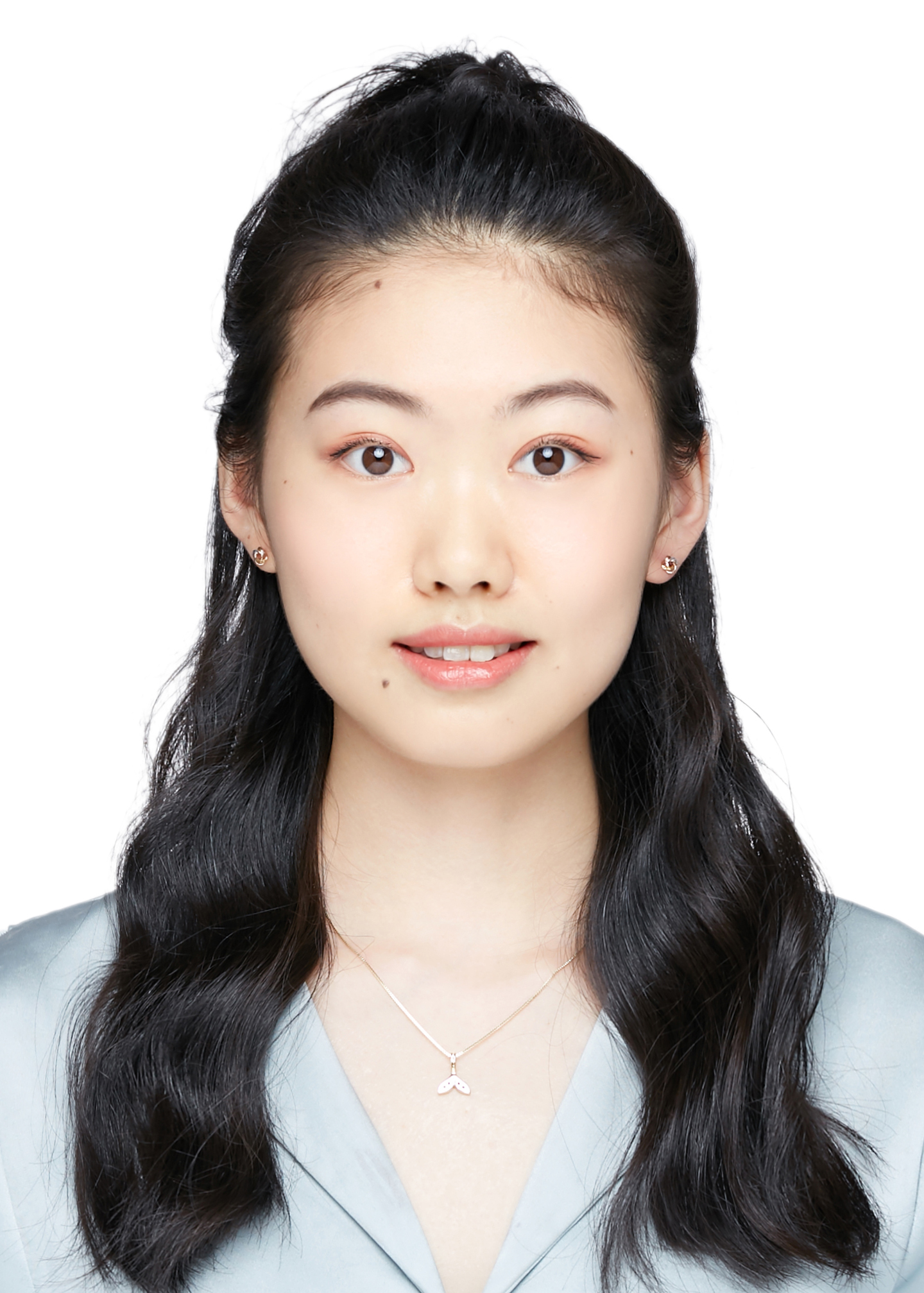}}]
	{Wenrong Chen}(S'19) received the B.E. degree in communication engineering, along with the Certification of the Talent Program in Yingcai Honors College, from the University of Electronic Science and Technology of China (UESTC) in 2019. She is currently working toward the Ph.D. degree with the National Key Laboratory of Wireless Communications, UESTC. She is a visiting Ph.D. student under supervision of Prof. Rui Zhang at the National University of Singapore. Her research and study interests include THz communication, integrated sensing and communication, stochastic geometry, networking optimization.
\end{IEEEbiography}
\begin{IEEEbiography}[{\includegraphics[width=1in,height=1.25in,keepaspectratio]{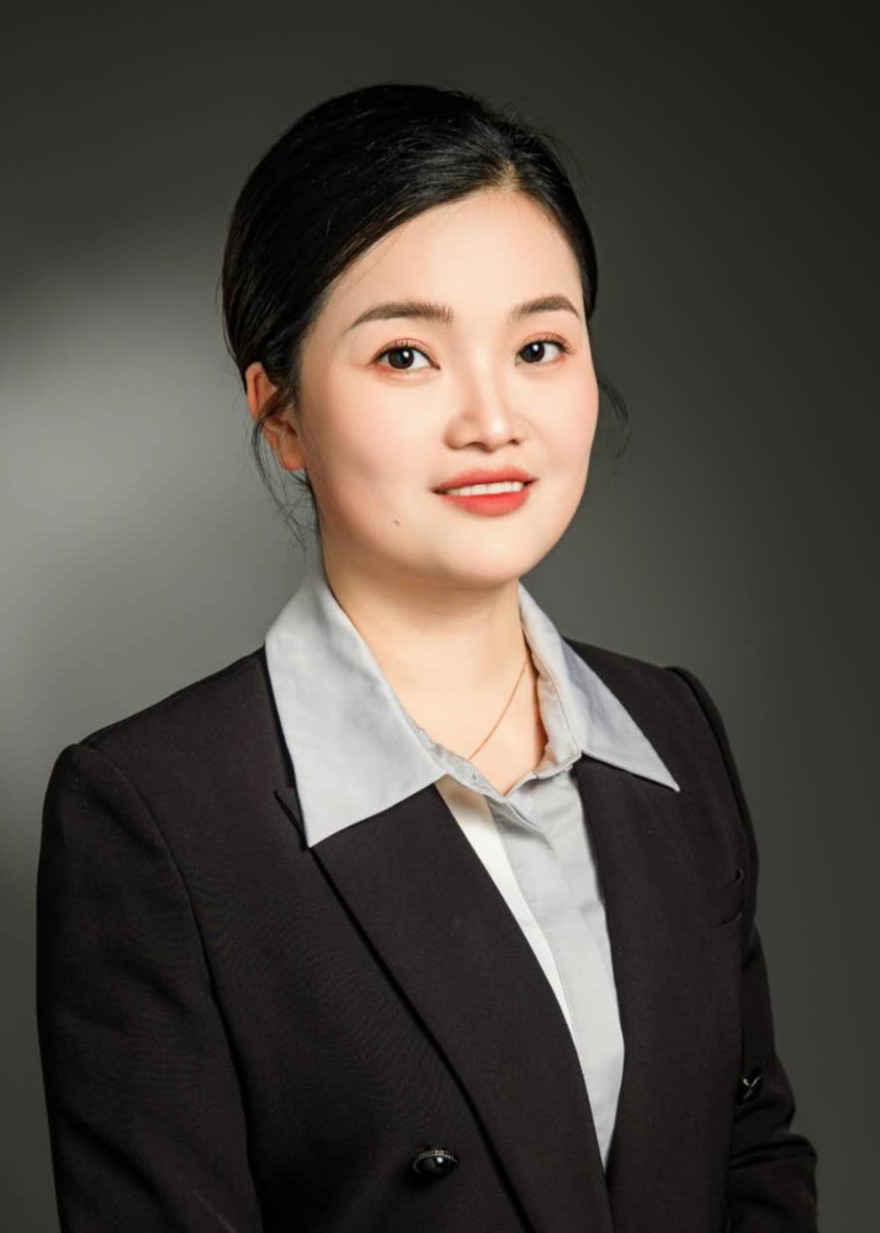}}]
	{Lingxiang Li}(S'13, M'17) received her M.S. and Ph.D. degrees in Electrical Engineering from University of Electronic Science and Technology of China (UESTC), Chengdu, China, in 2013 and 2017, respectively.
	She was a visiting Ph.D. student under supervisor of Prof. Athina P. Petropulu at Rutgers, The State University of New Jersey during 2015-2016,
	and a Postdoc Research Fellow collaborating with Prof. Tony Quek at Singapore University of Technology and Design (SUTD) during 2018.
	She is currently an Associate Professor with UESTC. Her research interests cover various aspects of signal processing and wireless communications, currently focusing on Terahertz communications, joint sensing and communications, and mobile computing.
\end{IEEEbiography}

\begin{IEEEbiography}[{\includegraphics[width=1in,height=1.25in,keepaspectratio]{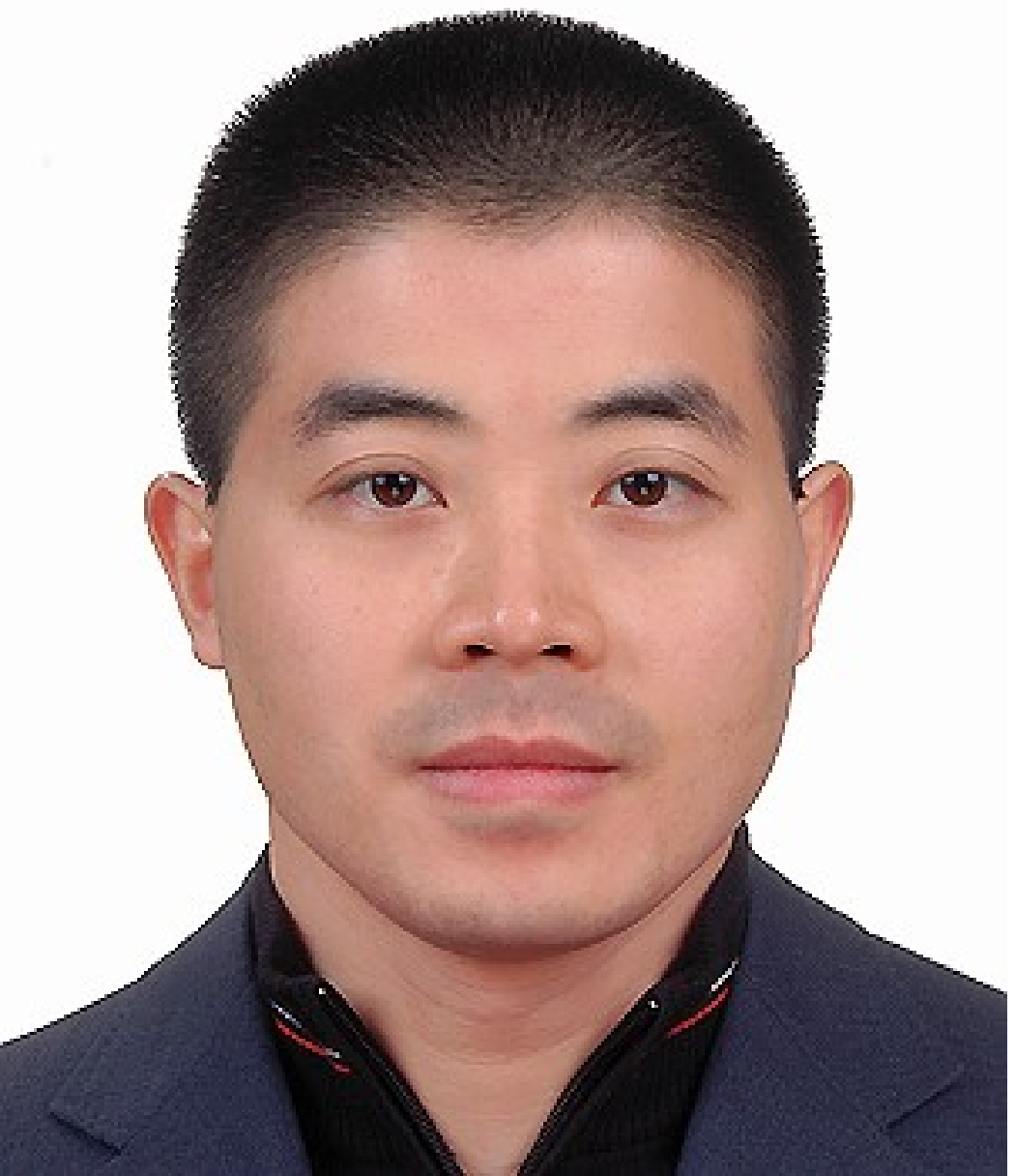}}]
	{Zhi Chen}(SM'16) received the B. Eng, M. Eng., and Ph.D. degrees in Electrical Engineering from University of Electronic Science and Technology of China (UESTC), in 1997, 2000, 2006, respectively. In April 2006, he joined the National Key Lab of Science and Technology on Communications (NCL), UESTC, and worked as professor in this lab from August 2013. He was a visiting scholar at University of California, Riverside during 2010-2011. His current research interests include 5G mobile communications, tactile internet, and Terahertz communication. Dr. Chen is a Senior Member of IEEE. He has served as a reviewer for various international journals and conferences, including IEEE Transactions on Vehicular Technology, IEEE Transactions on Signal Processing, etc.
\end{IEEEbiography}

\begin{IEEEbiography}[{\includegraphics[width=1in,height=1.25in,keepaspectratio]{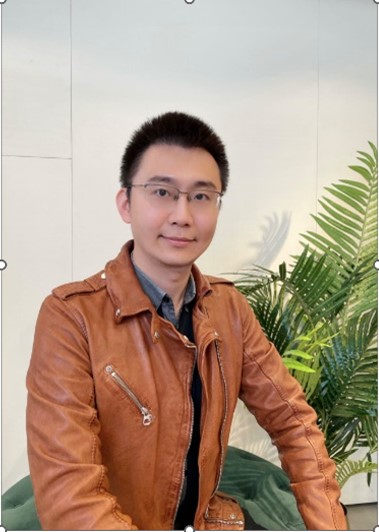}}]
	{Yuanwei Liu} (S'13-M'16-SM'19-F’24, \url{https://www.eee.hku.hk/~yuanwei/}) has been a (tenured) full Professor in Department of Electrical and Electronic Engineering (EEE) at The University of Hong Kong (HKU) since September, 2024. Prior to that, he was a Senior Lecturer (Associate Professor) (2021-2024) and a Lecturer (Assistant Professor) (2017- 2021) at Queen Mary University of London (QMUL), London, U.K, and a Postdoctoral Research Fellow (2016-2017) at King's College London (KCL), London, U.K. He received the Ph.D. degree from QMUL in 2016.  
	His research interests include non-orthogonal multiple access, reconfigurable intelligent surface, near field communications, integrated sensing and communications, and machine learning. 
	He is a Fellow of the IEEE, a Fellow of AAIA, a Web of Science Highly Cited Researcher, an IEEE Communication Society Distinguished Lecturer, an IEEE Vehicular Technology Society Distinguished Lecturer. He received IEEE ComSoc Outstanding Young Researcher Award for EMEA in 2020. He received the 2020 IEEE Signal Processing and Computing for Communications (SPCC) Technical Committee Early Achievement Award, IEEE Communication Theory Technical Committee (CTTC) 2021 Early Achievement Award. He received IEEE ComSoc Outstanding Nominee for Best Young Professionals Award in 2021. He is the co-recipient of the 2024 IEEE Communications Society Heinrich Hertz Award, the Best Student Paper Award in IEEE VTC2022-Fall, the Best Paper Award in ISWCS 2022, the 2022 IEEE SPCC-TC Best Paper Award, the 2023 IEEE ICCT Best Paper Award, and the 2023 IEEE ISAP Best Emerging Technologies Paper Award. 
	He serves as the Co-Editor-in-Chief of IEEE ComSoc TC Newsletter, an Area Editor of IEEE Communications Letters, an Editor of IEEE Communications Surveys \& Tutorials, IEEE Transactions on Wireless Communications, IEEE Transactions on Vehicular Technology, IEEE Transactions on Network Science and Engineering, IEEE Transactions on Cognitive Communications and Networking, and IEEE Transactions on Communications (2018-2023). 
	He serves as the (leading) Guest Editor for Proceedings of the IEEE on Next Generation Multiple Access, IEEE JSAC on Next Generation Multiple Access, IEEE JSTSP on Intelligent Signal Processing and Learning for Next Generation Multiple Access, and IEEE Network on Next Generation Multiple Access for 6G. 
\end{IEEEbiography}
\begin{IEEEbiography}[{\includegraphics[width=1in,height=1.25in,keepaspectratio]{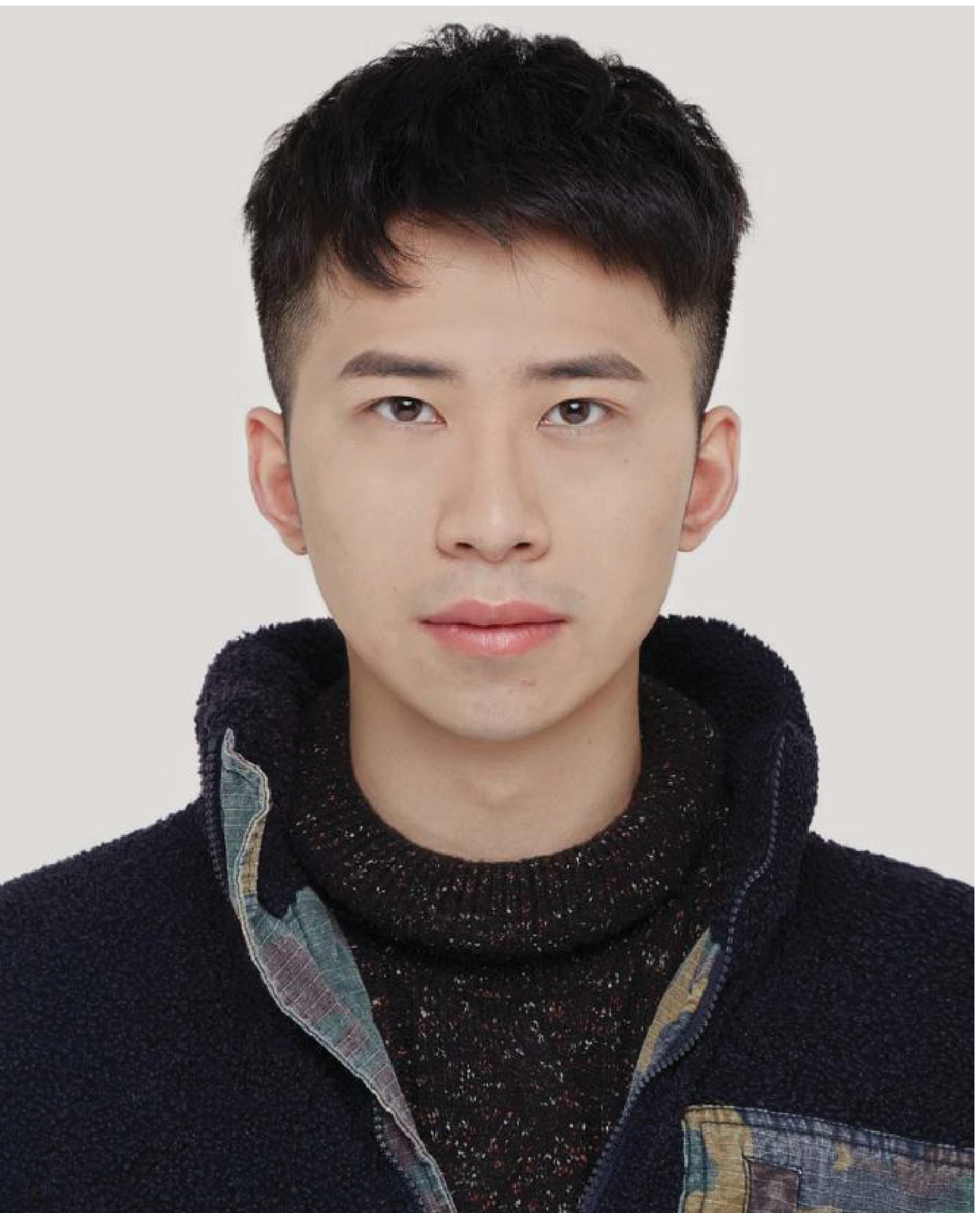}}]
	{Boyu Ning} received the B.E. degree in Communication Engineering, along with the Certification of the Talent Program in Yingcai Honors College, from the University of Electronic Science and Technology of China (UESTC), Chengdu, China, in 2018. He won the Most Comprehensive Scientific Research Award of the Oxford Study Programme from the Oxford University as a visiting student, in 2018. He was a recipient of the “Tang Lixin” Scholarship, in 2019. He is currently pursuing the Ph.D. degree with the National Key Laboratory of Science and Technology on Communications, UESTC. His research interests include Terahertz communication, intelligent reflecting surface, massive MIMO, physical-layer security, and convex optimization.
\end{IEEEbiography}
\begin{IEEEbiography}[{\includegraphics[width=1in,height=1.25in,keepaspectratio]{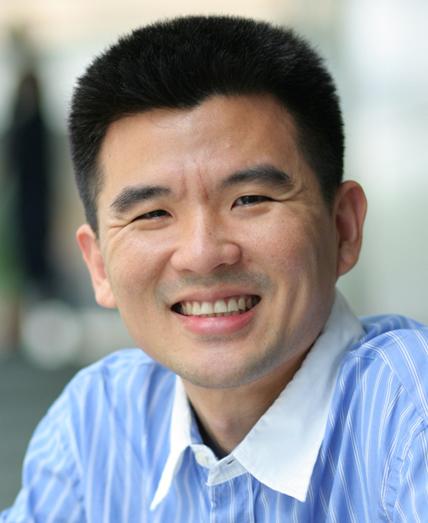}}]
	{Tony Q.S. Quek}(S'98-M'08-SM'12-F'18) received the B.E.\ and M.E.\ degrees in electrical and electronics engineering from the Tokyo Institute of Technology in 1998 and 2000, respectively, and the Ph.D.\ degree in electrical engineering and computer science from the Massachusetts Institute of Technology in 2008. Currently, he is the Cheng Tsang Man Chair Professor with Singapore University of Technology and Design (SUTD) and ST Engineering Distinguished Professor. He also serves as the Director of the Future Communications R\&D Programme, the Head of ISTD Pillar, and the Deputy Director of the SUTD-ZJU IDEA. His current research topics include wireless communications and networking, network intelligence, non-terrestrial networks, open radio access network, and 6G.
	
	Dr.\ Quek has been actively involved in organizing and chairing sessions, and has served as a member of the Technical Program Committee as well as symposium chairs in a number of international conferences. He is currently serving as an Area Editor for the {\scshape IEEE Transactions on Wireless Communications}. 
	
	Dr.\ Quek was honored with the 2008 Philip Yeo Prize for Outstanding Achievement in Research, the 2012 IEEE William R. Bennett Prize, the 2015 SUTD Outstanding Education Awards -- Excellence in Research, the 2016 IEEE Signal Processing Society Young Author Best Paper Award, the 2017 CTTC Early Achievement Award, the 2017 IEEE ComSoc AP Outstanding Paper Award, the 2020 IEEE Communications Society Young Author Best Paper Award, the 2020 IEEE Stephen O. Rice Prize, the 2020 Nokia Visiting Professor, and the 2022 IEEE Signal Processing Society Best Paper Award. He is an IEEE Fellow, a WWRF Fellow, and a Fellow of the Academy of Engineering Singapore.\end{IEEEbiography}

\end{document}